\documentclass[11pt]{book}
\usepackage[T1]{fontenc}
\usepackage[latin1]{inputenc}
\usepackage{babel}
\begin{document}

\begin{center}

{ {\bf STUDIES ON A NEW COSMOLOGICAL MODEL 

BASED ON COMPLEX METRIC }}

{ A thesis 

submitted for the award of }

{\sf Doctor of Philosophy

in 

Physics

\vspace{.25cm}
by}

\vspace{.15cm}
{ {\bf Moncy V. John 

\vspace {.15in}

Supervisor: Professor K. Babu Joseph }}

\vspace{.15 cm}

{ Department of Physics, Cochin University of Science and Technology

Kochi, Kerala 689641, India}

June 1999

\end{center}

\begin{center}

{{\bf Abstract}}

\end{center}

In this thesis, the implications of a new cosmological model are studied, which
has features similar to that of decaying vacuum cosmologies.  Decaying vacuum
(or cosmological constant $\Lambda $) models are the results of attempts to
resolve the problems that plague the standard hot big bang model in cosmology -
the problems which elude a satisfactory solution even after the two decades of
the advent of inflationary models, the first and much publicised cure to them.
We arrive at the present model by a radically new route, which extends the idea
of a possible signature change in the metric, a widely discussed speculation in
the current literature.  An alternative approach uses some dimensional
considerations in line with quantum cosmology and gives an almost identical
model.  Both derivations involve some fundamental issues in general theory of
relativity.  The model has a coasting evolution (i.e., $a\propto t$). It claims
the absence of all the aforementioned puzzles in the standard model and has
very good predictions for several measurable quantities.  In the first two
chapters of the thesis, we review the general theory of relativity, the standard
model in cosmology, its successes, the problems in it and also the most
successful of those attempts to solve them, namely, the inflationary and
decaying vacuum models.  In the third chapter, we present and discuss the new
cosmological model in detail.  The fourth chapter is concerned with quantum
cosmology.  We briefly review the canonical quantisation programme of solving
the Wheeler-DeWitt equation,  apply the procedure to our model and show that
it satisfies many of the much sought-after ideals of this formalism.  The last
chapter of the thesis discusses the solution of Einstein equations in the new
model in comparison with other ones, its connection with other coasting models,
the appearance of a Casimir type negative energy density in it and also the
prospects and challenges ahead for the model.

\newpage

\pagenumbering{roman} 
\begin{center} 
{\Large {\bf Acknowledgments}}

\end{center}

There are several people to whom I would like to express my
gratitude. Foremost is Professor K. Babu Joseph, my supervisor and mentor
who accepted me to his fold and kept me going all through these
years. This work was made possible only with the help of his vast
knowledge and unparalleled wisdom. On a personal note of grief,
homage to my friend and philosopher late Dr. Muralidharan M., who did
not wait to listen to this "truth game". A special thanks goes to my
friend and fellow traveler Sajith for his infinite patience with
computers and me. 

I am indebted to the Department of Physics, Cochin
University of Science and Technology where this work was carried out
and all the faculty members, students and office staff have been very
considerate to me. Especially, Professor M. Sabir, Professor V. C.
Kuriakose, Dr. Titus K. Mathew, Dr. K.P. Satheesh, Dr. Vinod, and Dr.
Suresh  were really encouraging. My friends Sivakumar, Ganapathy,
Taji, Aldrin, Vinoj, Ravikumar, Shaju, Jayadevan, Minu and others 
in the department were helpful in every matters. 

I owe
very much to the Management, Principal, Head of the Department of
Physics, all my colleagues and the nonteaching staff at the St. Thomas
College, Kozhencherri, my home institution, for all the freedom and
encouragement they have given me,
 which was beyond all expectations. A list of friends
and well wishers would be voluminous but it would be too inappropriate
not to mention the names of Dr. M.T. Simon, Professor N. Samuel Thomas,
Rev. Dr. Philip Varughese, Dr. P. J. Philip, Professor A. K. Kunnilethu and
Professor K. Thulasidharan Nair. Radhesh helped me with preparing the
graphs. 

The warmth and affection with which IUCAA (Inter University Center for Astronomy
and Astrophysics, Pune) welcomed me is unforgettable.  The encouragement by
Professor Jayant Narlikar, Professor Naresh Dadhich, Professor T.  Padmanabhan,
Professor Varun Sahni and the helping hand extended by Professor Ajit Kembhavi
have instilled much confidence in me.

I am grateful to my family which moulded me.
Those who  missed the most by my endeavours
are my wife Shirly, my daughter Shruthi and my son Arun. A special
wish goes to them. All those who supported me in this pursuit,
thank you.

\medskip

\newpage
\vspace{3in}

{\bf I dedicate this work to my father and  mother.}

\newpage

\begin{center}
{\bf {\Large Preface}}

\end{center}

Einstein's general theory of relativity (GTR) is perhaps the
profoundest theory concerning the physical world with regard to its
revolutionary content and the highly sophisticated mathematical apparatus
necessitated by it. While most theories of nature evolved as part of
experimental and observational encounters with physical situations by
innumerable scientists through generations, this theory, in its
complete form was conceived almost single handedly by this
intellectual giant and was much ahead of its time. Prospects of
putting it to direct test may ever remain poor, but the theory
assumes a central role in interpreting astrophysical and cosmological
data. In fact, the perspective of mankind on the cosmos was carried
to unforeseen heights in so short a period in this century mainly due
to GTR.

On the other hand, cosmology has never enjoyed the same status as
physics or astronomy till recent times, partly due to its speculative
nature and partly due to the lack of adequate observational data. But
during the past decade, with the launching of `Hubble Space
Telescope' and the `Cosmic Background Explorer' (COBE) satellite, a
wealth of information is pouring in from the deep skies. But since we
cannot experiment with the cosmos, one can only resort to
model-making and then to check how far the observational data agree
with the predictions of the model.  The most successful cosmological
model, with the least amount of speculatory inputs and maximum
consistency with observational facts is considered to be the
`standard' or the `hot big bang' model. The model predicts an early
hot phase for the universe, the relic of which is the cosmic
background radiation. In addition to background radiation, it predicts the
Hubble expansion and also the observed abundance of the light nuclei
in the universe.

However, there are certain problems in this picture, which are
identified and given serious attention in the past few years. Some of
these are directly dependent upon the simplifying assumptions taken
and some of them arise while trying to incorporate the ideas of particle
physics theories into the standard model. But there are problems like
the singularity, horizon, flatness and cosmological constant problems
which exhibit genuine inconsistencies in the model and require
substantial modifications in it. One of the most widely discussed such
modifications to standard model is the `inflation', which brings in
the possibility of an exponential expansion of the universe in its
early evolution, caused by the potential energy of a scalar field.
This scenario can successfully handle many of the problems, but does
not solve the singularity and cosmological constant problems and also brings
in a new `age' problem. Recently, some alternative cosmological models
have gained considerable attention in the literature under the title
`decaying-$\lambda $ cosmologies'. They have  a time-varying cosmological
constant, which helps to solve also the cosmological constant
problem, in addition to those ones inflation can solve.

In this thesis, we study the implications of a new cosmological
model, which has  features similar to those of   decaying-$\lambda $
cosmologies.  Apart from the presence of a time-varying cosmological
constant, the model has an evolution and thermal history quite close
to that of the standard model. At the same time, it claims the
absence of all outstanding problems in that model and has
very good predictions for several measurable quantities. We arrive at
this model by extending the idea of a possible signature change' in
the early universe, a widely discussed speculation which involves
some basic issues in the GTR. This extension leaves us in an
unphysical universe, but we have noticed that a proper interpretation
of the theory will enable us to obtain an excellent cosmological
model, with the essential features as summarised above. For the
purpose of comparison, we begin the thesis by introducing the
developments in the field of cosmology, starting from the
fundamentals. In the first two preliminary chapters, we review the
GTR, the standard model in cosmology, its successes, the problems in
it and also the most successful of those attempts to solve these
problems, namely, inflation and decaying vacuum models. Following
this, we present the new cosmological model  in Ch. 3 and discuss
its important features like thermal evolution, avoidance of
cosmological problems, prediction of observable quantities, etc..

The fourth chapter is devoted to quantum cosmology. Quantum
cosmology is the result of attempts to reconcile GTR and quantum
mechanics, the other major breakthrough in physics during this
century. This subject, which is still in its infancy, has a more direct
bearing on the conceptual foundations of physics. One route to this
goal is to write the wave equation for the universe (Wheeler-DeWitt
equation). We briefly review the achievements  in this
direction and then apply the procedure to the new cosmological model.
It is shown that the programme works exceedingly well in the new
context.

The concluding chapter of the thesis presents a discussion of the new
model in comparison with the other cosmological models.

Except in one subsection, we use natural units (in which $\hbar = c =
k_{B} = 1 $) throughout in the derivations. But when explicit
calculations are made, we convert the final results into conventional
units with the help of a table. Notations, sign conventions etc. are
adopted  mostly the same as that in \cite {landau}. Specifically, we
use Latin indices  $i,j,.. = 0,1,2,3$ and Greek indices $\mu, \nu
..=1,2,3.$

\tableofcontents

\newpage

\vspace{3in}

.... These games do not compliment or contradict cognitive reason. 
Therefore it is clear how they cannot be worked.  It cannot be an 
ontological reconciliation in which there are several aspects 
of being nor an epistemological one which assumes several types 
of knowledge.
Working this terrain pursues a fore-sight, a horizon of
expectations; not definitely of one goal to  truth, not even of
many roads to one truth, but perhaps, of many roads to many truths,
some to nowhere.

\medskip

Muralidharan M. in

\medskip

{\sl A Study of the Social and Ideological Implications of 

the Student - Teacher Discourses in the Upanishads}

Ph. D. Thesis, University of Calicut, 1993

\newpage
\pagenumbering{arabic}

\chapter{Relativistic Cosmology}

$\ddot{O}$pic in 1922 measured the distance to the Andromeda nebula to
be nearly equal to 450 Kpc, which when compared to the measured
radius $\approx 8$ Kpc of our own Milky-way is enormous. This was
conclusive proof of the fact that those observed spiral nebulae like
that of Andromeda are in fact island universes (galaxies), with a
size comparable to that of the Milky-way galaxy. Also it was the first
believable evidence that the universe extends to scales well above
that of our galaxy. The emergence of modern observational cosmology,
with the notion of galaxies as basic entities distributed over space,
can be traced back to this event. Around the same time, Slipher has
measured the spectral displacement of forty-one nearby galaxies and
thirty-six amongst them showed redshift.  In 1929 Hubble, on the
basis of Slipher's observations, proposed a linear relation - Hubble
law - between the distances to galaxies and their redshifts.
 The next landmark in observational cosmology was the discovery of
the cosmic microwave background radiation (CMBR) by Penzias and
Wilson in 1965.  Detailed observations on these three phenomena
\cite{peebles}, namely, distribution of galaxies, variation of galaxy
redshifts with distance and CMBR still remain the pillars of
observational cosmology.

Clearly, these observations require interpretations for any progress
to be made. The best thing one can do is to make a model by
extrapolating tested theories to the realm of cosmology and compare
the predictions of the model with more detailed observations.
However, this procedure involves certain judicious choices and
assumptions. At the range of scales involved, gravity is the only
known interaction to be counted and the most refined and tested
theory of gravity is  Einstein's general theory of relativity (GTR)
\cite{landau}-\cite{paddybook}. We discuss only models which use GTR
or some slight variants of it and hence a very brief review of this
theory is presented in Sec. \ref{sec-gtr}. Again, the application of
GTR to cosmology requires some simplifying assumptions for any
predictions to be made. First of all, we assume the cosmological
principle to be valid; i.e., at any given cosmic time, the
distribution of galaxies in the universe is assumed to be homogeneous
and isotropic at sufficiently large scales and also that the mean
rest frame of galaxies agrees with this definition of simultaneity.
In Sec. \ref{sec-homoiso}, we review models of the universe obeying
the cosmological principle, with different models having different
matter content. The last section in this chapter is devoted to a
brief review of the most popular, standard hot big bang model. We
explain how the model accounts for the observed facts at large, for
the benefit of comparison with the new cosmological model to be
presented in this thesis.

\section{General Theory of Relativity}
\label{sec-gtr}

The conventional route to GTR is to start from the observed
phenomenon of the equality of gravitational and inertial masses of
objects and then to elevate this equality to the `principle of
equivalence'. But this theory, which is primarily a geometric theory
- in the sense that gravitational field can be represented by the
metric tensor and freely falling bodies move along geodesics - can be
deduced also from an action principle.  For our purpose of
introducing a new cosmological model based on a complex metric, it is
convenient to adopt the latter approach. We first derive, by varying
an action, the equations of motion and the field equations in GTR,
making explicit the form of the energy-momentum tensor for various
types of matter. Then we make use of the opportunity to introduce
Einstein's famous cosmological constant, as it plays an important part
in our subsequent discussions. Lastly, by using the 3+1 split of
spacetime, it is described how to identify a suitable Lagrangian
density in this case, so as to enable writing the field equations as
Euler-Lagrange equations.

\subsection {Field Equations}

GTR is a theory of gravity which follows by requiring that the action
\cite{landau}, \cite{weinbook}-\cite{paddybook}

\begin{equation}
I = \frac {-1}{16\pi G}\int   \; R(g_{ik})\; \sqrt {-g}\; d^{4}x + 
 \int\;  \Lambda \; \sqrt {-g} \; d^{4}x
\equiv I_{G}+I_{M}   \label{eq:totact}
\end{equation}
be stationary under variation of the dynamical variables in it. $I$
is called the Einstein-Hilbert action. The first integral is  the
gravitational action $I_{G}$ where $R(g_{ik})$ is the curvature
scalar, $g_{ik}$  are the covariant components of the metric tensor
of the 4-dimensional spacetime, defined by the expression for the
line element

\begin{equation}
ds^{2} = g_{ik}dx^{i} dx^{k} \label{eq:le}
\end{equation}
and $g\equiv \det(g_{ik})$. $R(g_{ik})$ is given by

\begin{equation}
R= g^{ik} R_{ik}, \label{eq:r}
\end{equation}
where the $g^{ik}$ are the contravariant components of the metric
tensor and $R_{ik}$ is the Ricci tensor

\begin{equation}
R_{ik} = g^{lm}R_{limk} = R^{l}_{ilk}. \label{eq:rik}
\end{equation}
In the above, $R^{l}_{ilk}$ is the contracted form of the Riemann tensor

\begin{equation}
R^{l}_{imk} = \frac {\partial \Gamma ^{l}_{ik}}{\partial x^{m}} -
\frac  {\partial \Gamma ^{l}_{im}}{\partial x^{k}} + \Gamma ^{l}_{nm}
\Gamma ^{n}_{ik} - \Gamma ^{l}_{nk} \Gamma ^{n}_{im}
\end{equation}
and lastly, the Christoffel symbols $\Gamma ^{\; l}_{ik}\;, $  in terms of
the metric tensor are defined as

\begin{equation}
\Gamma ^{l}_{ik} = \frac {1}{2} g^{lm} \left( \frac {\partial
g_{mi}}{\partial x^{k}} + \frac {\partial g_{mk}}{\partial x^{i}} -
\frac {\partial g_{ik}}{\partial x^{m}} \right) .
\end{equation}

In the second integral in Eq. (\ref{eq:totact}), which is the matter action
$I_{M}$, $\Lambda $ corresponds to the matter fields present.
A general expression for $\Lambda $ is of the form

\begin{equation}
\Lambda = \Lambda (\phi ^{A}, \phi ^{A}_{,i}, x^{i}), \label{eq:lambda}
\end{equation}
where $\phi ^{A}$ $(A=1,2,3..)$ are a series of functions of
spacetime coordinates $x^{i}$ and ``$,i $" refers to differentiation with
respect to $x^{i}$. For example, the electromagnetic field
should have

\begin{equation}
\Lambda _{em} = -\frac{1}{16\pi }F_{ik}F^{ik} ; \qquad F_{ik} =
A_{k,i}- A_{i,k}. \label{eq:lambdaem}
\end{equation}
Here, $A_{i}$ are the scalar and vector potentials. For the scalar
field $\phi $ which appears in particle physics theories,

\begin{equation}
\Lambda _{\phi } = \frac {1}{2} g^{ik} \frac {\partial \phi}{\partial
x^{i}} \frac {\partial \phi }{\partial x^{k}} - V(\phi ),
\label{eq:lambdaphi}
\end{equation}
where $V(\phi )$ is the potential of the field. But for matter in the
form of particles, $I_{M}$ is written in a form different from  that
in Eq. (\ref{eq:totact}). As an example, consider particles
interacting with an electromagnetic field. We write the matter action
for the system as

\begin{equation}
I_{M,particles} = -\sum _{a} m_{a} \int ds_{a} - \sum _{a} e_{a}
\int \; A_{i}\; dx^{i} +  \int \; 
\Lambda _{em} \; \sqrt {-g}\; d^{4}x, \label{eq:imparticles}
\end{equation}
where $m_{a}$ is the mass,  $e_{a}$ the charge and the summation is
over all the particles $a$.  If  the action $I$ in (\ref{eq:totact})
is minimized by varying only the position of the worldline of  a
typical particle, keeping its endpoints fixed, we get the equation of
motion of the particle in the combined gravitational and other fields
with which it interacts. In the above example, the required equations
of motion for the particle are

\begin{equation}
\frac {d^{2}x^{i}}{ds_{a}^{2}} + \Gamma ^{i}_{kl} \frac
{dx^{k}}{ds_{a}} \frac {dx^{l}}{ds_{a}} = \frac {e_{a}}{m_{a}}
F^{\; i}_{l} \frac {dx^{l}}{ds_{a}}.
\end{equation}

On the other hand, the equations of motion for the fields,  i.e., the
field equations are obtained when we minimize the action $I$  by
varying only  the fields. For example, if we minimize the action with
$I_{M}$ given by equation (\ref{eq:imparticles}) by varying  $A_{i}$,
the Maxwell equations for the electromagnetic field are obtained:

\begin{equation}
F^{ik}_{;i} = 4\pi j^{k}.
\end{equation}
Here ``$ ;i $" refers to covariant differentiation with respect to
$x^{i}$.  In fact, the form (\ref{eq:lambdaem}) for $\Lambda _{em}$
was chosen in such a way that we obtain this result.

Lastly, the Einstein field equations, i.e., the equations of
motion for the gravitational field can be obtained by minimising the
action $I$ by varying the metric tensor $g_{ik}$. Note that this is
the only variation which will affect $I_{G}$. It can be seen that 
under the variation $g_{ik} \rightarrow g_{ik} + \delta g_{ik}$,

\begin{equation}
\delta I_{G} \equiv \frac {1}{16\pi G} \int \;  (R^{ik} -
\frac {1}{2} R\; g^{ik})\; \delta g_{ik} \; \sqrt {-g}\; d^{4}x. 
\label{eq:deltaig}
\end{equation}
The variation in the 
the matter action $I_{M}$ can be written as 

\begin{equation}
\delta I_{M} \equiv -\frac {1}{2}\int \; T^{ik}\; 
\delta g_{ik}\; \sqrt {-g}\; d^{4}x. \label{eq:deltaim}
\end{equation}
When $\Lambda $ in (\ref{eq:totact}) is of the general form
(\ref{eq:lambda}), $T^{ik}$, the energy-momentum tensor can be seen
to be of the form

\begin{equation}
T^{ik} = 2 \left[ \frac {1}{\sqrt {-g}} \left( \frac {\partial \Lambda \sqrt
{-g}}{\partial g_{ik,l}} \right) _{,l} - \frac {\partial \Lambda
}{\partial g_{ik}} - \frac {1}{2}\Lambda g^{ik} \right] . \label{eq:tik}
\end{equation}
For $\Lambda = \Lambda _{em}$ as in (\ref{eq:lambdaem}), this gives

\begin{equation}
T^{ik}_{em} = \frac{1}{4\pi }(\frac {1}{4} F_{mn}F^{mn} g^{ik} -
F^{i}_{l}F^{lk}). 
\end{equation}
For $\Lambda = \Lambda _{\phi }$ as in (\ref{eq:lambdaphi}), 
(\ref{eq:tik}) gives

\begin{equation}
T^{ik}_{\phi } = g^{il}g^{km} \frac {\partial \phi }{\partial x^{l}}
\frac {\partial \phi }{\partial x^{m}} -g^{ik} \Lambda _{\phi }.
\label{eq:tikphi} 
\end{equation}
For matter in the form of particles, as in the case of (\ref{eq:imparticles}),
 with the four-momentum $p^{i}_{a}$ and the energy of the particle $E_{a}$,

\begin{equation}
T^{ik}_{particles} = \sum _{a} \frac {1}{E_{a}} p^{i}_{a}p^{k}_{a}
\delta ^{3} (x-x_{a}).
\end{equation}
For a perfect fluid,  i.e., fluid having at each point a velocity
vector ${\bf v}$ such that an observer moving with this velocity sees
the fluid around him as isotropic, the above energy-momentum tensor can be
cast in the form

\begin{equation}
T^{ik}_{perfect\; fluid} = (p+\rho ) U^{i}U^{k} - p g^{ik} \label{eq:tikpf}
\end{equation}
where $U^{i} \equiv dx^{i}/ds$.

Combining (\ref{eq:deltaig}) with (\ref{eq:deltaim}) and 
putting $\delta I = \delta I_{G} +\delta
I_{M} =0$, we get the Einstein field equations as 

\begin{equation}
G^{ik} \equiv R^{ik}- \frac{1}{2} g^{ik}R = 8\pi G T^{ik}. \label{eq:gik}
\end{equation}
The Einstein equations also imply the energy conservation law

\begin{equation}
T^{i}_{k;l} =0. \label{eq:cons}
\end{equation}

\subsection {Cosmological Constant}

It is now instructive to see how $I_{G}$ is chosen in the form as in
(\ref{eq:totact}) \cite{landau}. As usual in writing variational principles, the
action shall be expressed in terms of a scalar integral $\int {\cal
G}\sqrt {-g}\; d^{4}x$, taken over all space and over the time
coordinate $x^{0} =t$ between two given values. Since the attempt is
to describe the gravitational field in terms of $g_{ik}$, which are
thus the `potentials', we shall require that the resulting equations
of the gravitational fields must contain derivatives of $g_{ik}$ no
higher than the second order.  For this, ${\cal G}$ should contain
only $g_{ik}$ and its first derivatives.  But it is not possible to
construct an invariant ${\cal G}$ (under coordinate transformations)
using $g_{ik}$ and the Christoffel symbols $\Gamma ^{i}_{kl}$ (which
contain only first derivatives of $g_{ik}$) alone, since both
$g_{ik}$ and $\Gamma ^{i}_{kl}$ can be made equal to zero at a given
point by appropriate coordinate transformations. Thus we choose $R$
in place of ${\cal G}$, though $R$ contains second derivatives of
$g_{ik}$. This is sufficient since the second derivatives in $R$ are
linear and the integral $\int R\; \sqrt {-g}\; d^{4}x$ can be written
as the sum of two terms: (1) an expression not containing the second
derivatives of $g_{ik}$ and (2) the integral of an expression in the
form of a four-divergence of a certain quantity.  By using 
Gauss's theorem, the latter can be transformed into an integral over
a hypersurface surrounding the four-volume over which the
integrations are performed. When we vary the action, the variations
of the second term vanish since by the principle of least action, the
variation of the field $g_{ik}$  at the limits of the region of
integration are zero. Thus $\int R\; \sqrt {-g}\; d^{4}x$ can
function as the gravitational action $I_{G}$.

However, as noted by Einstein himself, one can modify $I_{G}$ as

\begin{equation}
I_{G} = \frac {-1}{16\pi G}\int \; (R+2\lambda )\; \sqrt {-g}\; d^{4}x
\end{equation}
without violating the requirements on the action as described above,
where $\lambda $ is some new constant. Einstein used a 
very small $\lambda $ to obtain a stationary
universe. This constant is known as the
`cosmological constant' since when it is small, it will not
significantly affect the solutions, except in a cosmological context.
When Einstein came to know about the observational evidence
for the expansion of the universe, he decided to do away with it and 
described it as `the greatest
mistake in his life'. But this term $\lambda $ is one of the most
intriguing factors in current theoretical physics. It was later
recognised that  $\lambda $ can also be a function of
$x^{i}$ \cite{adler-ot}.

With the introduction of $\lambda (x^{i})$, the Einstein equation
(\ref{eq:gik}) can be written as 

\begin{equation}
R^{ik} -\frac{1}{2}Rg^{ik} -\lambda (x^{i})g^{ik} = 8\pi G T^{ik}.
\end{equation}
In view of its application in cosmology, the $\lambda $-term is
usually taken to the right hand side of this equation, after making a
substitution 

$$
\rho _{\lambda } = \frac {\lambda }{8\pi G},
$$
so that

\begin{equation}
R^{ik} - \frac {1}{2} R\; g^{ik} = 8\pi G (T^{ik} +
\rho _{\lambda } g^{ik}). \label{eq:giklambda}
\end{equation}
Using Eq. ({\ref{eq:tikpf}), one can see that the term $\rho
_{\lambda }g^{ik}$ in the above equation is identical to the
energy-momentum tensor for a perfect fluid having density $\rho
_{\lambda }$ and pressure $p_{\lambda } = -\rho _{\lambda }$.

\subsection {Lagrangian Density}

In the above subsection, we have seen that since the Ricci scalar $R$
contains second derivatives of $g_{ik}$ with respect to spacetime
coordinates,  the action  will contain second derivatives.  But
in fact, an alternative expression for $R$, which does not contain
any second derivatives of $g_{ik}$ can be found \cite{kolbturner}
(and references therein) using the
Arnowitt-Deser-Misner (ADM) 3+1 split of spacetime as

\begin{equation}
 R= K^{2}-K_{\mu \nu }K^{\mu \nu } - ^{3}\! R.
\end{equation}
This differs from the earlier expression (\ref{eq:r}) for $R$  by a
possible four-divergence. In the present case, we have conceived  a
foliation of spacetime into space-like hypersurfaces $\Sigma _{t}$
labeled by $t$, which is some global time-like  variable. $^{3}R$ is the
scalar curvature of this 3-dimensional surface,  $K_{\mu \nu }$ are
the components of the extrinsic curvature of $\Sigma _{t}$ defined by

\begin{equation}
K_{\mu \nu } = \frac {1}{2N} \left( N_{\mu \mid \kappa } + N_{ \kappa
\mid \mu }- \frac {\partial h_{\mu \nu }}{\partial t}\right)
\end{equation}
and

\begin{equation} 
K = h^{\mu \nu }K_{\mu \nu}.
\end{equation}
$N^{\mu }$ is called the shift vector,  $N$, the lapse function and
$h_{\mu \nu }= n_{\mu }n_{\nu }-g_{\mu \nu}$ (where $n_{\mu }$ is the
vector field normal to $\Sigma _{t}$) is the metric induced on this
3-space with $\sqrt {-g} = N\sqrt {h}$. " $\mid $ " denotes covariant
differentiation with respect to the spatial metric $h_{\mu \nu }$.
The line element (\ref{eq:le}), in terms of the lapse $N$ and shift
$N^{\mu }$ is given as

\begin{equation} 
ds^{2}= g_{ik}dx^{i}dx^{k} = (N\; dt)^{2} - h_{\mu \nu} (N^{\mu }dt +
dx^{\mu })(N^{\nu }dt+dx^{\nu })
\end{equation}
so that

\begin{equation}
g_{ik}=
\left[
\begin{array}{clcr}
N^{2}-N_{\mu }N_{ \nu }h^{\mu \nu } & -N_{\nu }  \\
-N_{\mu } & -h_{\mu \nu }   \\
\end{array}
\right] \label{eq:cometric}
\end{equation}
and

\begin{equation}
g^{ik}=
\left[
\begin{array}{clcr}
\frac {1}{N^{2}} & -\frac {N^{\nu }}{N^{2}}  \\
-\frac {N^{\mu }}{N^{2}} & \frac {N^{\mu } N^{\nu }}{N^{2}}-h^{\mu \nu }   \\
\end{array}
\right] . \label{eq:contrametric}
\end{equation}

Thus the Lagrangian density to be used in the gravitational action
$I_{G}$ is 

\begin{equation} 
{\cal L}_{G}= -\sqrt {-g}R/16\pi G = -\frac {1}{16\pi G}\sqrt {h} N \left(
K^{2}-K_{\mu \nu }K^{\mu \nu }-^{3}\! R\right) .\label{eq:lg}
\end{equation}

The changes corresponding  to that in the metric tensor are to be
implemented in the  matter action too. For example, in the case of a scalar
field, Eq. (\ref{eq:cometric}) and (\ref{eq:contrametric}) are to be used in the matter Lagrangian
density 

\begin{equation} 
{\cal L}_{\phi }= \sqrt {-g}\Lambda _{\phi }=\sqrt {-g}\left[ \frac
{1}{2}g^{ik}\frac {\partial \phi }{\partial x^{i}}\frac {\partial
\phi }{\partial x^{k}}-V({\phi )}\right] . \label{eq:lphi}
\end{equation}

One can write the Euler-Lagrange equations corresponding to
variations with respect to $N^{\mu }$, $N$ and other dynamic
variables in the total Lagrangian density ${\cal L}_{G}+ {\cal
L}_{M}={\cal L}$.  ($N^{\mu }$ and $N$ are not dynamical variables;
their time derivatives do not appear in ${\cal L}$. In fact, these are
Lagrange multipliers so that after the variation one can fix some
convenient gauge for them.) The equations obtained by varying with
respect to $N$ and $N^{\mu }$ are `constraint equations' and they
contain only first derivatives.  Variation with respect to the other
dynamical variables leads to field equations. The resulting equations
can be seen to be the same as those obtained from the Einstein field 
equations (\ref{eq:gik}). We shall make this explicit using specific
examples in the next section.

\section {Homogeneous and Isotropic Cosmologies}
\label{sec-homoiso}

This section serves two purposes. First, it illustrates the formalism
of GTR summarised in the last section by applying it to cosmology.
But more importantly, it introduces the general framework of  models
which obey the cosmological principle \cite{landau}-\cite{paddybook}.
Friedmann models form the basis of the standard hot big bang model
whereas models with a minimally coupled scalar field paves the way
for the inflationary cosmological models.  We obtain the field
equations for these models in the conventional way, but in the last
subsection, we demonstrate their derivation using the Euler-Lagrange
equations.

\subsection {Friedmann Models}

If the distribution  of matter in space is homogeneous and isotropic,
we can describe the spacetime by the maximally, spatially symmetric
Robertson-Walker (RW) metric and obtain an important class of
solutions to the Einstein field equations that are  of much
significance in cosmology. The RW line element is given by

\begin{equation}
ds^{2} = dt^{2} - a^{2}(t) \left[ \frac {dr^{2}}{1-kr^{2}} +
r^{2} (d\theta ^{2} + \sin ^{2}\theta\; d\phi ^{2})\right] .
\label{eq:rwle} 
\end{equation}
$a(t)$ is the scale factor of the spatial expansion and $k=0$, $+1$ or
$-1$ which, in the respective order, corresponds to flat, positively
curved or negatively curved spacelike hypersurfaces of constant $t$.
Let us apply the  formalism of GTR to this simple case.

Evaluating $R$ and $R_{ik}$ using (\ref{eq:r}) and (\ref{eq:rik}),
the Einstein equations (\ref{eq:gik})  can be written  for the perfect
fluid described by (\ref{eq:tikpf}) in a comoving frame with $U^{i} =
(1,0,0,0)$, which describes a homogeneous and isotropic distribution
of matter as

\begin{equation}
\frac {\dot {a}^{2}}{a^{2}} +\frac {k}{a^{2}} = \frac {8\pi
G}{3} \rho, \label{eq:t-t}
\end{equation}

\begin{equation}
2 \frac {\ddot{a}}{a} + \frac {\dot{a}^{2}}{a^{2}} +\frac {k}{a^{2}} =
 -8\pi G p. \label{eq:s-s}
\end{equation}
Differentiating (\ref{eq:t-t}) and combining with (\ref{eq:s-s}) gives

\begin{equation}
\frac {d(\rho a^{3})}{da} +3pa^{2} =0 \label{eq:cons1}
\end{equation}
or equivalently

\begin{equation}
\dot {\rho} = -3 \frac {\dot {a}}{a} (\rho  +p), \label{eq:cons2}
\end{equation}
which is the conservation law for energy-momentum (\ref{eq:cons}) in
this case.  Combining (\ref{eq:t-t}) and (\ref{eq:s-s}) in a
different way, we get another useful result

\begin{equation}
\frac {\ddot {a}}{a} = -\frac {4\pi G}{3}(\rho  +3p). \label{eq:ddota}
\end{equation}

The solutions of these equations require, however, some additional
information in the form of an `equation of state' relating $\rho $ and
$p$. In most commonly encountered problems, we can write this
relation as 

\begin{equation}
p=w\rho. \label{eq:es}
\end{equation}
It can be shown that for extreme relativistic matter, $w=1/3$ and for
nonrelativistic matter (dust), we have $w=0$. These equations
(\ref{eq:t-t})-(\ref{eq:es}) were first obtained and studied by A.
Friedmann and models based upon these are usually called Friedmann
models. They predict either an expanding or contracting
universe.

Eq. (\ref{eq:cons1}) can immediately be solved to obtain

\begin{equation}
\rho \propto a^{-3(1+w)}. \label{eq:rhoa}
\end{equation}
If there are more than one noninteracting component in $\rho $
that are separately conserved,  (\ref{eq:cons1}) and hence (\ref{eq:rhoa})
 are applicable to each. For relativistic matter, the density $\rho _{m,r}
\propto a^{-4}$  and for nonrelativistic matter, $\rho _{m,nr} \propto
a^{-3}$.  The variation of $\rho $ with $a$ for other values of $w$ can also
be deduced from (\ref{eq:rhoa}); for $w=-1/3$, $\rho \propto a^{-2}$ and for
$w=-1$, $\rho $ is a constant.

To study the variation of $a $ with $t$, we  make a few
definitions. The quantity 

\begin{equation}
H(t) \equiv \frac {\dot {a}}{a} \label{eq:h}
\end{equation}
is called the Hubble parameter which measures the rate of expansion of
the universe. The deceleration parameter $q(t)$ is defined through
the relation

\begin{equation}
\frac {\ddot {a}}{a} \equiv -q(t) H^{2}(t) \label{eq:q}
\end{equation}
and the critical density  as 

\begin{equation}
\rho _{c} \equiv \frac {3}{8\pi G} H^{2}.
\end{equation}
Another important quantity is the density parameter

\begin{equation}
\Omega (t) = \frac {\rho }{\rho _{c}}.
\end{equation}
Using these definitions, Eq. (\ref{eq:t-t}) can be written as

\begin{equation}
\Omega -1 = \frac {k}{a^{2}H^{2}} . \label{eq:o-1}
\end{equation}
The $k=0$ case is a special one where $\Omega =1$ or $ \rho = \rho _{c}$.
 Using (\ref{eq:rhoa}) in (\ref{eq:t-t}) gives the solution in this case as 

\begin{equation}
a(t) \propto t^{2/3(1+w)}. \label{eq:a}
\end{equation}
For the $k=+1$ case, $\Omega >1$, $q>1/2$ and the universe expands to
a maximum and then recollapses.  For $k=-1$, $\Omega <1$, $q<1/2$ and
it expands for ever. The $k=0$ case is critical in the sense that it
just manages to expand for ever.

\subsubsection {de Sitter Models}

Instead of matter, if the RW spacetime contained only a cosmological
constant, Eq. (\ref{eq:giklambda}) (with $T^{ik}=0$) leads to 

\begin{equation}
\frac {\dot {a}^{2}}{a^{2}} + \frac {k}{a^{2}} 
 = \frac {8\pi G}{3} \rho _{\lambda }, \label{eq:t-tdS}
\end{equation}

\begin{equation}
2\frac {\ddot {a}}{a} + \frac {\dot {a}^{2}}{a^{2}} + \frac
{k}{a^{2}}  = 8\pi G \rho _{\lambda }. \label{eq:s-sdS}
\end{equation}
The field equations are thus similar to a Friedmann model with
equation of state $p_{\lambda }=-\rho _{\lambda }$; i.e., with
$w=-1$.  Thus a positive (negative) $\rho _{\lambda }$ has a
repulsive (attractive) effect so that we have an accelerating
(decelerating) cosmic evolution with $\ddot {a}>0$ ($\ddot {a}<0$).
It is the repulsive force due to a constant positive $\rho _{\lambda
}$, which Einstein made use of in his stationary universe model to
prevent it from collapsing due to other matter distributions present.

Equations (\ref{eq:t-tdS}) and (\ref{eq:s-sdS}) are particularly
simple to solve in the flat case with $k=0$. The solution, with
$H\equiv (8\pi G\rho _{\lambda }/3)^{1/2}$ = constant, is obtained as 

\begin{equation}
a(t) \propto e^{Ht} \label{eq:exp}
\end{equation}
If we define

\begin{equation}
H= \sqrt {\frac {8\pi G\rho _{\lambda }}{3}} \; \tanh ^{k} \left(
\sqrt {\frac {8\pi G \rho _{\lambda }}{3}} t \right) ,
\end{equation}
a solution can be found  also for the $k=\pm 1$ cases. For $k=+1$,

\begin{equation}
a(t) \propto  H^{-1} \; \cosh Ht
\end{equation}
and for $k=-1$,

\begin{equation}
a(t) \propto H^{-1} \; \sinh Ht.
\end{equation}
The model with positive $\rho _{\lambda }$ is called the de Sitter
model, after  W. de Sitter, who solved it for the first
time. The model with  $\rho _{\lambda }$ negative, is called the
anti-de Sitter model.

\subsection{Models With a Scalar Field}

Another special case of interest is that of a RW spacetime filled
with a minimally coupled scalar field $\phi $, whose energy-momentum
tensor is given by (\ref{eq:tikphi}). With the assumption that $\phi
$ is spatially homogeneous and depends only on time, Einstein
equations can be written in a similar manner as that in
(\ref{eq:t-t})-(\ref{eq:s-s}) 

\begin{equation}
\frac {\dot {a}^{2}}{a^{2}} + \frac {k}{a^{2}} = 
 \frac {8\pi G}{3} \left[ \frac {\dot {\phi }^{2}}{2} + V(\phi
)\right] ,\label{eq:t-tphi}
\end{equation}

\begin{equation}
2\frac {\ddot {a}}{a} + \frac {\dot {a}^{2}}{a^{2}} + \frac
{k}{a^{2}} = - 8\pi G 
\left[ \frac {\dot {\phi }^{2}}{2} - V(\phi )\right] . \label{eq:s-sphi}
\end{equation}

The equation of motion for $\phi $ can be obtained by using the
conservation law for energy-momentum (\ref{eq:cons}) as

\begin{equation}
\ddot {\phi }+ 3 \frac {\dot {a}}{a}
\dot {\phi } + \frac {dV(\phi ) }{d\phi }=0. \label{eq:consphi}
\end{equation}
It shall be noted that when the field is displaced from the minimum
of its potential and when  $\dot {\phi }^{2} \ll V(\phi )$, Eq.
(\ref{eq:t-tphi}) and (\ref{eq:s-sphi}) are similar to the Einstein
equations (\ref{eq:t-tdS}) and (\ref{eq:s-sdS}), written for the
spacetime containing only a cosmological constant (where we identify
$V(\phi ) = \rho _{\lambda }$).  In this context, $\rho _{\lambda }$
is usually called the vacuum energy density.  The solutions for
spacetimes which contain a cosmological constant in addition to
matter were studied by G. Lemaitre and such models are generally
referred to as Friedmann-Lemaitre-Robertson-Walker (FLRW)
cosmologies.

\subsection {Field Equations as Euler-Lagrange Equations}

Lastly, let us demonstrate how the Einstein equations in different
models are obtained as Euler-Lagrange equations under the variation
of the action. For the RW spacetime, under the ADM 3+1 split,
$N^{\mu}=0$, $K_{\mu \nu }= -(1/N) (\dot {a}/a)h_{\mu \nu }$,
$^{3}\! R=6k/a^{2}$, 

\begin{equation}
R=\frac {6}{N^{2}} \frac {\dot {a}^{2}}{a^{2}}- \frac {6k}{a^{2}}
\label{eq:ricci} 
\end{equation}
 and the action, using (\ref{eq:lg}) and (\ref{eq:lphi}), is

\begin{eqnarray}
I & = & I_{G}+I_{M}  =  \int ({\cal L}_{G}+{\cal L}_{M})d^{4}x \nonumber \\
& = & \int N\sqrt {h}
\left[ -\frac
{1}{16\pi G}\left( \frac {6}{N^{2}}\frac {\dot {a}^{2}}{a^{2}} - \frac
{6k}{a^{2}} \right) + \left( \frac {\dot {\phi }^{2}}{2N^{2}}-V(\phi
)\right)\right] d^{4}x. \label{eq:iel}
\end{eqnarray}
Integrating the space part, we get

\begin{eqnarray}
I & = & 2\pi ^{2}\int Na^{3}\left[ -\frac {1}{16\pi G}\left( \frac
{6}{N^{2}}\frac {\dot {a}^{2}}{a^{2}} - \frac {6k}{a^{2}} \right) +
 \frac {\dot {\phi
}^{2}}{2N^{2}} - V(\phi )\right] dt  \nonumber \\
& \equiv & \int L\; dt. \label{eq:ielphi}
\end{eqnarray}
Using the Lagrangian $L$, we may write the Euler-Lagrange equations
for the variables
$N$, $a $ and $\phi $  and fixing the
gauge $N=1$ to obtain the same  Einstein equations
(\ref{eq:t-tphi})-(\ref{eq:consphi}). 

Similarly, for a de Sitter model which contains only a cosmological
constant, the Lagrangian can be taken to be

\begin{equation}
L=2\pi ^{2} Na^{3}\left[ -\frac {1}{16\pi G}\left( \frac
{6}{N^{2}}\frac {\dot {a}^{2}}{a^{2}}- \frac {6k}{a^{2}} \right) -
 \rho _{\lambda }\right] \label{eq:iellambda}
\end{equation}
The Einstein equations (\ref{eq:t-tdS}) and (\ref{eq:s-sdS}) are
obtained on writing the Euler-Lagrange equations corresponding to
variations with respect to $N$ and $a$, in the gauge $N=1$.

\section {The Standard Model - Its Successes}
\label{sec-stand}

The standard model \cite{peebles}-\cite{paddybook} claims to have
the least amount of speculatory inputs into cosmology, while having
maximum agreement with observations. It is based upon the following
assumptions:  (1) At  the very large scales of the size greater than
clusters of clusters of galaxies, the universe is homogeneous and
isotropic and hence is describable by the RW metric and (2) It is
filled with relativistic/ nonrelativistic matter. Then the
fundamental equations governing the evolution of the universe are
those obtained earlier (\ref{eq:t-t})-(\ref{eq:es}) with $w=0$ or
$1/3$. These models predict an expanding or contracting universe and
belong to Friedmann cosmologies. We now discuss the three major
success stories of the model, juxtaposing them with the current
status of observational cosmology.

\subsection  {The Hubble Expansion}

 The 1929 discovery of a linear redshift-distance relation for
galaxies by Hubble, if interpreted as due to Doppler effect,
establishes the case for an expanding phase for the universe at
present and was a primary piece of evidence in support of the
standard model.  At present, the  expansion rate, characterised by
the Hubble parameter (\ref{eq:h}) is in the range $H_{p} = 100\;h$ Km
s$^{-1}$ Mpc$^{-1}$ ; $h =0.7\pm 0.05$. (The subscript $p$ refers to
the present epoch.) The Hubble radius
$H_{p}^{-1}\approx 0.9 \times 10^{28}h^{-1}$ cm $\approx 2.9 \times
10^{3} h^{-1}$ Mpc gives a measure of the size of the presently
observed universe. The deceleration parameter defined by (\ref{eq:q})
is estimated to be lying in the range $-0.5<q_{p}<2$. Also the
density parameter, as per current estimates is given by $0.1 \leq
\Omega _{p}\leq 2$. The age of the universe, measured by direct
observational dating techniques is  $t_{p}\approx 5\times
10^{17}$ s. Though these observations are not precise enough, they
however confirm the Hubble expansion of the universe.

The observed redshift $z$ of galaxies can be related to the  scale
factor $a$ as 

\begin{equation}
1+z = \frac {a(t_{p})}{a(t_{1})} \label{eq:z}
\end{equation}
where $t_{1}$ is the time at which the light is emitted. If we assume
that the universe contains both radiation and matter, according to
equation (\ref{eq:rhoa}), before some time $t_{eq}$ in its history, radiation
will dominate over matter. In the standard cosmology, $t_{eq}$ is
estimated to be $\approx 1.35 \times 10^{11} \Omega ^{-3/2}h^{-3}$ s.
For a universe with flat space sections (i.e., $k=0$), (\ref{eq:a})
gives 
$a\propto t^{1/2}$ for the relativistic era and $a\propto t^{2/3}$
for the nonrelativistic era. Assuming that the
changeover is instantaneous, we can write

\begin{equation}
a=At^{2/3}, \qquad t> t_{eq}\label{eq:a1}
\end{equation}

\begin{equation}
a=Bt^{1/2}, \qquad t<t_{eq}. \label{eq:a2}
\end{equation}
Matching the two relations at $t=t_{eq}$, one  estimates

\begin{equation}
\frac {B}{A} = \frac {t_{eq}^{2/3}}{t_{eq}^{1/2}} \approx 0.7\times
10^{2}\Omega ^{-1/4}h^{-1/2} {\hbox {s}}^{1/6}. \label{eq:b/a}
\end{equation}
This value will be of use in evaluating expressions of the type
(\ref{eq:z}) in the standard flat models.
In both the other cases with $k=\pm 1$, 
 we can regard the universe as
nearly flat when $a$ was smaller than $a_{p}$ by a few orders of
magnitude (See flatness problem: Sec. \ref{sec-problems}).

\subsection {Cosmic Microwave Background Radiation}

Another important milestone in the development of the standard model
was the discovery  of the cosmic microwave background radiation
(CMBR) by Penzias and Wilson in 1965. The spectrum of CMBR is
consistent with that of a blackbody at temperature $T_{p}\approx 2.73
K$. It endorses the view that there was a more contracted state for
the universe, which ought to have been denser and hotter than the
present. According to the standard model, the universe cools as it
expands and when the temperature reaches $T\approx 4000 K$, matter
ceases to be ionised, the electrons join the atoms. Radiation is then
no more in thermal equilibrium with matter (matter-radiation
decoupling) and the opacity of the radiation drops sharply. The
radiation we see now as CMBR is conceived as the relic of that last
scattered  at the time of decoupling. In fact, the CMBR was predicted
by Gamow  in 1948 and its discovery, perhaps, is the strongest
observational evidence in support of the standard model.

We can derive an expression for the total relativistic matter
(radiation)  density
$\rho _{m,r}$ in terms of temperature by the following
argument \cite{kolbturner}. (We 
use conventional units in this subsection.) For an
ideal gas, there are $1/h^{3}$ number of states located in unit
volume of $\mu $-space, where $h$ is the Planck's constant. The
number of states in volume $V$ with momentum less than $P$ will be
$(4/3) \pi P^{3} V/h^{3}$. The occupancy of a single state is

$$
 \frac {1}{ e^{(E_{A}(P) -\mu _{A})/kT_{A}} \pm 1}.
$$
$+(-)$ signs correspond to Fermi (Bose) statistics, $\mu _{A}$ is the
chemical potential and  $T_{A}$ is the temperature of the species $A$
which is assumed to be in equilibrium and $E_{A}(P) =
(P^{2}c^{2}+m^{2} c^{4})^{1/2}$, the energy of a particle in the
species $A$. Then the number of particles of type $A$ with momentum
between $P$ and $P+dP$ per unit volume of space is

\begin{equation}
n_{A}(P)dP = \frac {g_{A}}{2\pi ^{2}\hbar ^{3}} \frac {P^{2}dP}
{e^{(E_{A}(P) -\mu _{A})/kT_{A}} \pm 1},
\end{equation}
where $g_{A}$ is the number of spin degrees of freedom. In the
extreme relativistic  $(T_{A} \gg m_{A})$ and nondegenerate
$(T_{A}\gg \mu _{A})$ limit, the energy density, which corresponds to
species $A$ is

\begin{eqnarray}
\rho _{A} = \int _{0}^{\infty} E_{A}(P)n_{A}(P)dP & = & g_{A} \sigma
T_{A}^{4}  \qquad {\hbox {(Bosons)}} \\
& = & (7/8) g_{A} \sigma T_{A}^{4} \qquad {\hbox {(Fermions)}} 
\end{eqnarray}
where $\sigma = \pi ^{2}k^{4}/30\hbar ^{3}c^{3} 
= 3.782 \times 10^{-15}$ erg m$^{-3}$ K$^{-4}$. The total energy density 
contributed by all the relativistic species together can be written
as 

\begin{equation}
\rho _{m,r}c^{2}  = g_{tot} \sigma T^{4}, \label{eq:rhor}
\end{equation}
where 

\begin{equation}
g_{tot} = \sum _{(A=Bosons)}g_{A} (T_{A}/T)^{4} + \sum _{(A=Fermions)}
(7/8)g_{A}(T_{A}/T)^{4}
\end{equation}
is the effective number of spin degrees of freedom at temperature T.
In the very early universe, $g_{tot}$ is evaluated to be nearly
equal to 100.

The expression for $\rho _{m,r}$ as given by (\ref{eq:rhor}) is a
reasonable speculation if we agree to look upon the CMBR as the relic
of a hot early universe. To obtain another useful result in the study
of the thermal history of an expanding universe, we  apply the
second law of thermodynamics, in its familiar form, to a physical
volume $V=a^{3}$;

\begin{equation}
kT\;dS = dE +pdV = d(\rho c^{2}a^{3}) + pd(a^{3})
\end{equation}
and also use (\ref{eq:cons1}), which is a statement of the first law of
thermodynamics. It is easy to see that

\begin{equation}
\frac {dS}{dt} = \frac {1}{kT} \left[ \frac {d}{dt}(\rho
c^{2}a^{3})+p\frac {d}{dt}(a^{3})\right] = 0.
\end{equation}
This implies that the entropy per comoving volume element of unit
coordinate volume $V=a^{3}$, under thermal equilibrium, is a
constant. i.e., 

\begin{equation}
 S= \frac {(\rho c^{2} +p)}{kT} a^{3} = {\hbox {constant}}. \label{eq:s}
\end{equation}

Thus in the standard model, the universe expands adiabatically. Eq. 
(\ref{eq:rhor}) implies that for radiation with $\rho _{m,r}
\propto a^{-4}$, $aT$ is a constant. In the relativistic era, for a
$k=0$ universe, this may be used to write

\begin{equation}
t= \left( \frac {3c^{2}}{32\pi G \sigma} \right)^{1/2} g_{tot}^{-1/2}
T^{-2}. 
\end{equation}
The times at which radiation reaches various temperatures can be evaluated
using this expression.

\subsection {Primordial Nucleosynthesis}

The third important success of the standard model is the prediction of
primordial nucleosynthesis \cite{weinbook}-\cite{paddybook}.
According to this theory, when the age of the universe was of the
order of 1 s, the temperature was of the order of $10^{10}$ K and the
conditions were right for nuclear reactions which ultimately led to
the synthesis of significant amounts of D, $^{3}$He, $^{4}$He and
$^{7}$Li. The yields of these light elements, according to the model,
depends on the baryon to photon ratio $\eta$ and the number of very
light particle species, usually quantified as the equivalent number
of light neutrino species $N_{\nu }$. The predictions of the
abundance of the above four light elements agree with the
observational data provided the free parameters $\eta $ and $N_{\nu }
$ in the theory have values in the range

\begin{equation}
2.5 \times 10^{-10}\leq \eta \leq 6\times 10^{-10}, \qquad N_{\nu }.
\leq 3.9
\end{equation}
In turn, if we accept the present abundance of light nuclei, the
density parameter for baryons $\Omega _{B}$ may be predicted from the
above to be  lying in the range 

\begin{equation}
0.01\leq \Omega _{B} \leq 0.15,
\end{equation}
which agrees with measured  values. Furthermore, the bounds on
$N_{\nu }$

\begin{equation}
N_{\nu } = 3.0 \pm 0.02
\end{equation}
agree with particle accelerator experiments.

\chapter {Problems and Solutions}

\section {The Standard Model - Problems}
\label{sec-problems}

The three major observational facts, namely, a linear
redshift-distance relation, a perfect blackbody distribution for CMBR
which corresponds to a more or less uniform temperature and the
observed abundance of light elements have clearly established a case
in favour of the standard, hot big bang model. However, this is only
a broad brush picture and there are several loose ends to be sorted
out when we go into details. There are issues like the formation of 
structures etc., which call for refinements of the theory.
But here we focus attention on another class of puzzles, usually
called `cosmological problems', which deserve special attention since
they indicate the possible existence of some inconsistencies in the
standard model and hence do require substantial modifications in its
underlying postulates. The most serious among them are the following.

\subsubsection {Singularity Problem}

The assumptions in the standard model (See Sec. \ref{sec-stand})
  are in tune with the validity of the strong energy
condition $\rho +3p \geq 0$ and $\rho +p \geq 0$.  This, when
combined with some topological assumptions and causality conditions
lead to strong singularity theorems which imply that a singularity,
where the geometry itself breaks down, is unavoidable. In the
cosmological context, this singularity corresponds to the instant of
creation, the big bang, where quantities like matter density,
temperature, etc., take unbounded values. The universe comes into
existence at this instant, violating the law of conservation of
energy, which is one of the most cherished principles of physics
\cite{narpad}. This is called the singularity problem.

\subsubsection {Flatness Problem}

From equation (\ref{eq:o-1}), which may be written in the form 

\begin{equation}
\Omega -1 = \frac {1}{\frac {8\pi G}{3} \frac {\rho a^{2}}{k} -1},
\label{eq:flat} 
\end{equation}
it is easy to see that for $\Omega $ being close to unity, $|\Omega
-1|$ grows as $a^{2}$ during the radiation dominated era ($\rho
\propto a^{-4}$) and as $a$ in the matter dominated era ($\rho
\propto a^{-3}$). Thus since $\Omega (t_{p})$ is still of the order of
unity, at early times it was equal to 1, to a very high precision. For
instance

$$
\Omega (10^{-43}\; {\hbox {s}}) = 1\pm O(10^{-57}),
$$

$$
\Omega (1\; {\hbox {s}}) = 1\pm O(10^{-16}).
$$
This means, for example, that if $\Omega $ at the Planck time
$t_{pl}=5.4\times 10^{-44}$ s  was slightly greater than 1, say $\Omega
(10^{-43)}$ s$ = 1+10^{-55}$, the universe would have collapsed millions
of years ago. The standard model cannot explain why the universe was
created with such fine-tuned closeness to $\Omega =1$ \cite{dicke}.
This is the flatness problem. 

\subsubsection {Horizon Problem}

The CMBR is known to be isotropic with a high degree of precision.
Two microwave or infrared antennas pointed in opposite directions in
the sky do collect thermal radiation with $\Delta T/T\leq 10^{-5}$,
$T$ being the black body temperature. In the context of the standard
model, this is puzzling since these two regions from which CMBR of
strikingly uniform temperature is emitted cannot have been in causal
contact at any time in the past \cite{horizon}. The problem can be
explicitly stated as follows. According to the standard model, the
proper distance to the horizon of the presently observed universe is
of the order of $H_{p}^{-1}$. Since distances scale as $a(t)$, at any
time in the past, say $t_{s}$, the size of the same part of the
universe was $[a(t_{s})/a(t_{p})] H_{p}^{-1}$. But the distance a
light signal can travel by the time $t_{s}$ is equal to the proper
distance to the horizon at that time; i.e.,

\begin{equation}
 d_{hor }(t_{s}) = a(t_{s})\int _{0}^{t_{s}}\frac { dt}{a(t)}.
\label{eq:hor} 
\end{equation}
If the presently observed part of the universe was to be in causal
contact at $t_{s}$, a necessary (though not sufficient) condition is 

\begin{equation}
d_{hor}(t_{s}) > \frac {a(t_{s})}{a(t_{p})} H_{p}^{-1}. \label{eq:hcond}
\end{equation}
The isotropy of the CMBR, which was traveling unobstructed since the
time of decoupling $(t_{dec})$, indicates that the presently observed
part of the universe was in causal contact at least by that time.
Hence, one would expect the above condition to be satisfied for some
time $t_{s} <t_{dec}$. In the standard model, $t_{dec} \approx
10^{13}$ s and the time at which the universe changes from
relativistic to nonrelativistic era is $t_{eq}\approx 10^{11}$ s.  
Using these and also some typical values $\Omega =1$, $h=3/4$ and
$t_{p}= 5\times 10^{17}$ s, Eqs. (\ref{eq:a1})-(\ref{eq:b/a}) will
help us to evaluate both sides of condition (\ref{eq:hcond}). It can be
seen that the right hand side of this condition is greater than the
left by a factor of $2.5 \times 10^{7}/t_{s}^{1/2}$ for $t_{s}
<t_{eq}$ and by a factor of $0.63\times
10^{6}/[(t_{eq}^{1/2}/40)+3t_{s}^{1/3} -3t_{eq}^{1/3}]$ for
$t_{s}>t_{eq}$, thus violating the condition. For $t_{s}=t_{eq}$,
this ratio is approximately equal to $80$ and for $t_{s}=t_{dec}$, it
is $\approx 10$.
This means that the presently observed part of the
universe was not even in causal contact at the time of decoupling.
Yet the surface of last scattering of radiation appears very much
isotropic. This is the horizon problem.

Further, for the successful prediction of the primordial
nucleosynthesis, the universe has to be homogeneous at least as early
as $\approx 1$ s. The condition (\ref{eq:hcond}) is then violated by
a very wide margin.

\subsubsection {Problem of Small Scale Inhomogeneity}

The assumption in the standard model that the universe is homogeneous
and isotropic is justifiable at least in the early epochs, before the
matter-radiation equality. The remarkably uniform temperature of
CMBR on all angular scales upto quadrupole, is ample evidence for this.
But recent measurements show anisotropies (of the order of $10^{-5}$
or so) in CMBR in a systematic way and these anisotropies directly
sample irregularities in the distribution of matter at the time of
last scattering. It is believed that once the universe becomes matter
dominated, small density inhomogeneities grow via the Jeans
instability. Density inhomogeneities are usually expressed in a
Fourier expansion

\begin{equation}
\frac {\delta \rho (\vec {x})}{\rho } = \frac {1}{(2\pi )^{3}} \int
\delta _{k}\exp (-i\vec {k}. \vec {x})d^{3}k,
\end{equation}
where $\rho $ is the mean density of the universe, $\vec {k}$ is the
comoving wave number associated with a given mode and $\delta _{k}$
is its amplitude. So long as a density perturbation is of small
magnitude(i.e., $\delta \rho /\rho \ll 1$), its physical wave number and
wave length scale with $a(t)$ as $k_{phys} = k/a(t)$, $\lambda
_{phys} = a(t)\times 2\pi /k$. Once a perturbation becomes nonlinear,
it separates from the general expansion and maintains an
approximately constant physical size. The inhomogeneity at present
is: stars ($\delta \rho /\rho \approx 10^{30}$), galaxies ($\delta
\rho /\rho \approx 10^{5}$), clusters of galaxies  ($\delta \rho /\rho 
\approx 10 - 10^{3}$), superclusters or clusters of clusters of
galaxies ($\delta \rho /\rho \approx 1$) and so on. Based upon this
fact that nonlinear structures exist today, and the fact that in the
linear regime fluctuations grow as $a(t)$ in the matter dominated
epoch, we can calculate the amplitude of perturbations that existed on
these scales at the epoch of decoupling. It should be possible to
account for the anisotropies in the  CMBR detected by the COBE
satellite on this basis. The problem with this scenario of small
scale inhomogeneity is that in the standard model, last scattering
occurred at a redshift of around 1000 with the Hubble radius at that
time subtending an angle of only around $1^{o}$, while CMBR shows
anisotropies on all angular  scales upto the quadrupole
\cite{kolbturner}. 

This problem is closely related to the horizon problem in that if one
imagines causal, microphysical processes acting during the earliest
moments of the universe and giving rise to primeval density
perturbations, the existence of particle horizons in the standard
cosmology precludes production of  inhomogeneities on the scales of
interest.

\subsubsection {Problem of the Size of the Universe}

If we follow the standard evolution, the size of the comoving volume
corresponding to the present Hubble volume at the Planck time
$t_{pl}$  can be evaluated using
(\ref{eq:a1})-(\ref{eq:b/a})  as $10^{-4}$ cm. This is much greater
than the Planck length $l_{pl}=1.616\times 10^{-33}$ cm, the only
natural length scale available.  This is the problem of the size of the
universe
\cite{lindebook}.

\subsubsection {Entropy Problem }

 Most of the entropy
in the universe exists in the form of relativistic matter. The
radiation entropy in the present Hubble volume may be evaluated
using (\ref{eq:rhor}) in (\ref{eq:s}) as 

\begin{equation}
S_{p} = \left( \frac {\rho _{m,r} +p_{r}}{T}\right) _{p} \frac
{4}{3}\pi (H_{p}^{-1})^{3} \approx 10^{88},
\end{equation}
with $g_{tot.}\approx 2$ and $T_{p} =2.73$ K. The entropy at $t_{pl}$,
again if we follow the standard evolution, will be the same as
$S_{p}$. Where do such large numbers come from, is the entropy
problem \cite{lindebook}.

\subsubsection {Monopole Problem}

This is another problem closely related to the horizon problem. The
Grand Unified Theories (GUTs) predict that as the universe cools down
and the temperature reaches a value $\approx 10^{28}$ K, a
spontaneous symmetry breaking occurs and as a result, magnetic
monopoles are copiously produced. However, no such monopoles have yet
been detected. This is the monopole problem \cite{brand}.

The monopoles which are expected to be produced are of mass $\approx
2\times 10^{-8}$ g. At the end of the GUT epoch $(t_{c})$, we expect at
least one monopole per horizon size sphere to be produced. The
horizon radius at $t_{c}$ is given by $2t_{c}$.
The radius of the same part of the universe at present is $d_{hor}(t_{c})
a(t_{p})/a(t_{c})$. Thus the present number density of monopoles will
be of order

\begin{equation}
n_{monopole} (t_{p}) \approx \frac {1}{\frac {4\pi }{3} \left[
d_{hor}(t_{c}) \frac {a(t_{p})}{a(t_{c})}\right] ^{3}}.\label{eq:mono}
\end{equation}
With $T(t_{c}) \approx 10^{28}$ K and $ d_{hor}(t_{c}) = 2t_{c}
\approx 10^{-26}$cm, we can evaluate (with $a\propto T^{-1}$ and
$T_{p} = 2.73$ K),

\begin{equation}
n_{monopole} (t_{p}) \approx 10^{-6} \hbox {cm}^{-3}.
\end{equation}
With the monopole mass $\approx  2\times 10^{-8}$ g, we get

\begin{equation}
\rho _{monopole}(t_{p}) \approx 2\times 10^{-14} {\hbox {g cm}}^{-3}.
\end{equation}
This is much greater than the closure density $\approx 10^{-29}$ g
cm$^{-3}$ of the present universe and if it were true, the universe
would have collapsed much earlier. If the horizon problem is solved
before $t_{c}$, the same dynamical mechanism would solve the monopole
problem too. Further, this problem is not related to cosmology alone;
 if particle physics turns out to be discarding the hypothesis
regarding monopole production,  this problem will also disappear.

\subsubsection{Cosmological Constant Problem}

If we assume that the universe contains a cosmological constant $\rho
_{\lambda } \equiv \lambda /8\pi G $ in addition to matter, then by
measuring the values of the Hubble parameter $H=\dot {a}/a$  and the
deceleration parameter $q=-\ddot {a}/aH^{2}$ and using the Einstein
equations, one can find the magnitude of $\rho _{\lambda }$. Current
estimates \cite{perl,kochanek} of this value is of the order of the
critical density $\approx  10^{-29} $ g cm$^{-3}$. If $\rho _{\lambda
}$ is viewed as arising from the potential energy of a scalar field
employed in field theoretic models, then no known symmetry principle
in quantum field theory requires that its value be so small like this
when compared to the Planck density $\rho _{pl}= 0.5 \times 10^{94}$
g cm$^{-3}$. That is, the measured value of $\rho _{\lambda }$ is
smaller than the expected value $\rho _{pl}$ by 122 orders of
magnitude. This is the cosmological constant problem
\cite{weinpap}.

\section {Attempts to Modify the Standard Model}
\label{sec-attempts}

Let us now extrapolate backward in time the standard evolution for a
universe with flat space using  $ \Omega = 1$ , $h=0.75$,
$t_{p} = 5\times 10^{17}$s and accepting the standard model values
$z_{dec}\approx 1000$, $z_{eq}\approx 13500$ and $T_{nuc}\approx
10^{10}$ K (the subscripts $dec$ and $nuc$ refers, respectively, to
the decoupling and nucleosynthesis epochs).  We also normalise $a(t)$
such that $a(t_{p})$ is  equal to unity.  Using the relation $1+z
=a(t_{p})/a(t)$, we get

\begin{equation}
a_{dec} \approx 10^{-3}.
\end{equation}
Writing $a =A t^{2/3}$ in the matter dominated era, we can evaluate
$t_{dec}\approx 1.58\times 10^{13}$ s. Similarly, with $a=Bt^{1/2}$
in the relativistic era, we get $a_{eq}=8.18\times 10^{-5}$ and
$t_{eq} =3.18\times 10^{11}$ s. Using $a\propto T^{-1}$ where $T$ is
the radiation temperature, we can evaluate $a_{nuc} = 3\times
10^{-10}$ and $t_{nuc} = 5.2$ s. If we extrapolate still backward
with the same expression $a=Bt^{1/2}$ till the Planck era, when the
energy density is $\rho _{pl} = 0.5\times 10^{94}$ g cm$^{-3}$, we
get $a(\rho = \rho _{pl}) = 2.155\times 10^{-32}$ and $t(\rho =\rho
_{pl})=2.7 \times 10^{-44}$ s $\approx t_{pl}$. As mentioned in Sec.
\ref{sec-problems}, this value of $a$ corresponds to $10^{-4}$ cm for
a region of size $10^{28}$ cm at present and is very large when
compared to Planck length.

The horizon problem and the problem of generation of density
perturbations above the present Hubble radius can be better
understood in this context. We have already stated that to account for
the remarkable isotropy of CMBR, the condition is (\ref{eq:hcond})
(with $a_{p}=1$): i.e.,

\begin{equation}
d_{hor}(t_{dec})  > a_{dec}H_{p}^{-1},
\end{equation}
where $d_{hor}(t_{dec})$ is given by equation (\ref{eq:hcond}). 
This statement may
be extended to explain the anisotropies which correspond to the
aforementioned density perturbations. Let us define 

\begin{equation}
d_{comm.}(t_{1}, t_{2}) = \int _{t_{1}}^{t_{2}} \frac {dt}{a(t)},
\label{eq:comm1} 
\end{equation}
which is the communication distance light has traveled between times
$t_{1}$ and $t_{2}$, evaluated at present. Then the condition for
generation of density perturbations above the Hubble radius can be
expressed as

\begin{equation}
d_{comm.}(t_{pl}, t_{dec}) > H_{p}^{-1}.\label{eq:comm2}
\end{equation}
This is identical to the condition (\ref{eq:hcond}), provided we
extend the lower limit of integration in (\ref{eq:hor}) to $t_{pl}$.
In the above scenario of standard evolution, the left hand side of
Eq. (\ref{eq:comm2}) may be evaluated as equal to $ 1.1\times
10^{27}$ cm whereas the right hand side is approximately $1.2\times
10^{28}$ cm.  This is an alternative way of stating the puzzles in
the standard model.

Based on the behaviour of the scale factor alone, it was recently
argued \cite{hu,liddle} that some nonstandard evolution is essential
for the solution of these problems. The argument is based upon the
understanding that the standard model gives a reliable and tested
accounting of the evolution of the universe, at least from the time
of nucleosynthesis onwards. Hence it was asserted that if we do not
want to jeopardise the successes of the standard model, the evolution
should be standard, as described above, from $t \approx 1$ s onwards.
Then one has to face the question of how to maximize (\ref{eq:comm1}),
so that (\ref{eq:comm2}) is satisfied. Those models which do not
violate the  condition $\rho +3p \geq 0$ has $\ddot
{a}\leq 0$. In this scenario, (\ref{eq:comm1}) can be maximized by
assuming a coasting evolution $\ddot {a} =0$ or $a\propto t$. More
specifically, Liddle \cite{liddle} assumes

\begin{equation}
a(t) = \frac {a_{nuc}}{t_{nuc}}t, \qquad a<a_{nuc}.
\end{equation}
In this picture, at the epoch when $\rho = \rho _{pl}$, we
have $t\approx 3.75\times 10^{-22}$ s. Between this epoch and that of
nucleosynthesis, the maximum possible communication distance will be
$\approx 3.55\times 10^{22}$ cm, which is only a small fraction of
$d_{comm} (t_{nuc},t_{dec}) \approx 10^{27}$ cm. Thus the above problems
cannot be solved in this picture. One has to conceive some
nonstandard evolution characterised by $\ddot {a}>0$ or $\rho +3p<0$.
But this will necessitate some drastic modification to the standard
model in that  the existence of some kind of energy density with
equation of state $p=w\rho$, $w<0$ should be accepted, at least in
the early epochs.  One such case is a universe filled with the
potential energy of a scalar field. The resulting evolution, which
resembles that of de Sitter cosmologies is called `inflation'.

\section {Inflation}
\label{sec-inflation}

It is well known how inflation solves  the cosmological problems
\cite{sahni}. In all inflationary models, the universe which emerges
from the Planck epoch, after a brief period of standard evolution (or
sometimes without it) finds itself containing the potential energy of
a scalar field, which is generally called the inflaton field. This
field is initially displaced from the minimum of its potential and it
rolls down slowly to that minimum. All viable inflationary models are
of this slow rollover type or can be recast as such. Then the
governing equations are (\ref{eq:t-tphi})-(\ref{eq:consphi}) with
$\dot {\phi }^{2}\ll V(\phi )$ so that during this phase, the
universe expands quasiexponentially as in (\ref{eq:exp}) with $H$
remaining a constant.  That the inclusion of a minimally coupled
scalar field will lead to such a dynamics was known much earlier. It
was  Guth \cite{guth} who showed that this phenomenon can possibly lead to the
solution of cosmological problems. He showed that this exponential
expansion stretches causally connected regions of size $H^{-1}$ by an
amount $\exp (H\Delta t)$ and consequently regions of size
$H^{-1}\approx l_{pl}$ reach a  size $H^{-1}\exp H\Delta t
\approx 10^{-4}$ cm by the time inflation ends, provided $H\Delta t
\approx 67$. This will help the universe to evolve as per the
standard model for the rest of the time. This also will resolve the
horizon problem.  From (\ref{eq:flat}), it is seen that during this
period when $\rho _{\phi }\equiv V(\phi )\approx $ constant, for a
closed universe,  $\Omega -1$ decreases as $a^{-2}$ and by the end of
the inflationary era, $\Omega $ can be arbitrarily close to unity.
Similar is the behaviour of an open universe.  Thus one need not
start with any fine tuned  initial conditions at the Planck epoch to
get a nearly flat universe at present. This solves the flatness
problem.

The inflationary stage gives way to the standard evolution when the
scalar field reaches its minimum. During this process, the entropy of
the universe increases enormously. This will solve the entropy
problem.  The monopole problem disappears along with the horizon
problem.

The above are features generic to all inflationary models. But for
the successful implementation of the mechanism, one has to decide on
what type of field to constitute the inflaton field, what the potential
of the field is and what initial conditions are to be specified.
There are numerous inflationary models which differ in these matters.
Guth \cite{guth} proposed his model as a possible solution to
horizon, flatness and monopole problems in which the grand unified
models tend to provide phase transitions that lead to an inflationary
scenario of the universe. A grand unified model begins with a simple
gauge group G which is a valid symmetry at the highest energies. As
the energy is lowered, the theory undergoes a hierarchy of
spontaneous symmetry breaking transitions into successive subgroups.
At high temperatures, the Higgs fields of any spontaneously broken
gauge theory would loose their expectation values, resulting in a
high temperature phase in which the full gauge symmetry is restored.
The effective potential $V(\phi, T)$ of the scalar field $\phi $ has
a deep local minimum at $\phi =0$, even at a very low temperature
$T$.  As a result, the universe remains in a supercooled vacuum state
$\phi =0$, which is a false vacuum, for a long time. The
energy-momentum tensor of such a state would be the same as that  in
equations (\ref{eq:t-tphi})-(\ref{eq:s-sphi}) with $\dot {\phi }^{2}
\gg V(\phi )$ and the universe expands exponentially until the false
vacuum decays. This phenomenon is termed  a first order phase
transition, which occurs at some critical temperature $T_{c}$. As the
universe cools through this temperature, one would expect bubbles of
the low temperature phase to nucleate and grow and these bubbles
contain the field $\phi _{0}$, which corresponds to the minimum of
the effective potential $V(\phi )$. The universe will cool as it
expands and it will then supercool in the high temperature phase.
When the phase transition finally takes place at this low
temperature, the latent heat is  released and the universe is
reheated to a temperature comparable to $T_{c} $. If the universe
supercools to 28 or more orders of magnitude, sufficient entropy will
be generated due to bubble wall collisions and thermalisation of
energy.

As pointed out by Guth himself, the major problem with this scenario
is that if the rate of bubble nucleation is greater than the speed of
expansion of the universe, then the phase transition occurs very
rapidly and inflation does not take place. On the other hand, if the
vacuum decay rate is small, then the bubbles cannot collide and the
universe becomes unacceptably inhomogeneous. 

In order to improve this scenario, Linde \cite{new1} and, Albrecht
and Steinhardt \cite{new2}
independently suggested the `new inflationary
model'. The crucial 
difference between the new and old inflationary models is  in the
choice of the effective potential $V(\phi ,T)$ and that the latter is
a second order spontaneous symmetry breaking phenomenon. The new
choice was the Coleman-Weinberg potential which has a bump near $\phi
=0$. For the decay from the false vacuum to the true vacuum, the
system has to tunnel across the bump and then it slowly rolls down the
potential. After reaching the minimum of the potential, it executes
damped oscillations, during which energy is thermalised and entropy is
increased. In the new inflation, a typical size of the bubble at the
moment of its creation is $\approx 10^{-20}$ cm. After the exponential
expansion, the bubble will have a size much greater than the
observable part of the universe, so that we see no inhomogeneities
caused by the wall collisions. The drawback of the new inflationary
model is that it requires fine tuned initial values for the field.

Whereas the `old' and `new' inflationary models are the result of
spontaneous symmetry breaking, the chaotic inflation \cite{chaos}
proposed by Linde does not contain any phase transition at all. The
scalar field is not part of any unified theory and its only purpose
is to implement inflation. The potential in chaotic inflation is
assumed to be of the simple form

\begin{equation}
V(\phi ) = \frac  {l } {n} \phi ^{n}.
\end{equation}
The minimum of the potential is at $\phi =0$ and so it has nothing to
do with spontaneous symmetry breaking. With 
sufficiently large initial value,  the field $\phi $  may roll slowly
so that $a(t)$ rapidly approaches the asymptotic regime

\begin{equation}
a(t) = a_{0} \exp \left[ \frac {4\pi }{n}\left( \phi _{0}^{2} -\phi
^{2}(t)\right) \right]. 
\end{equation}
Linde envisions that the initial distribution of $\phi_{0}$ is
chaotic', with different values in different regions of the
universe. In the $n=4$ case, to obtain sufficient inflation, say 60
e-folds, $\phi _{0}$ must be greater than about $4.4\; m_{pl}$ where
$m_{pl}$ is the Planck energy. The
model in its simplicity is not definite enough to discuss reheating. 

A major drawback of chaotic inflation is  the required smoothness
of the initial inflationary patch. For inflation to occur, we see the
condition

\begin{equation}
(\bigtriangledown \phi )^{2} \ll \frac {l }{4} \phi _{0}^{4}.
\end{equation}
If we take the dimension of the region over which $\phi $ varies by
the order of unity to be $L$, then

\begin{equation}
(\bigtriangledown \phi )^{2} \approx \left( \frac {\phi
_{0}^{2}}{L}\right) ^{2}    \ll \frac {l}{4} \phi _{0}^{4}
\end{equation}
which implies $L\gg (\phi _{0}/m_{pl})H^{-1}$. This means that, for
sufficient inflation, $\phi $ must be smooth on a scale much greater
than the Hubble radius, a condition which does not sound very chaotic.

In addition to these most widely discussed ones, there are many other
models which exhibit inflation, but having a variety of features
\cite{sahni} (and references therein). The 
`natural inflation' model is the one having the potential $V=V_{0}
\cos ^{2}(\phi /m)$. Those models named `power law inflation' have
either $a\propto t^{p}$, $p>1$ with potential $V(\phi ) = V_{0} \exp
(-\mu \phi )$ or $a\propto (t_{c} - t)^{-q}$, $q>1$ which obeys the
induced gravity action, a variant of GR. Other  models which make use
of non-Einstein theories of gravity like the latter one are
`Starobinski model' (higher derivative gravity), `Kaluza-Klein
inflation' (higher dimensional Kaluza-Klein theories), `extended
inflation' (Brans-Dicke theory), `pre-big bang inflation'
(superstring theories) etc.. This shows that even after the two
decades since the development of the theory of inflation, it still
lacks a unique model.

\subsubsection{Age and Other Problems}

The most important prediction of the inflationary  models is that the
universe is almost spatially flat with the present value of the
density parameter $\Omega _{p}$ very close to unity. This in turn
implies that the combination $H_{p}t_{p}\approx 2/3$. A major set
back to inflationary models was in fact this prediction. Recent
observations \cite{freedpi} put this
value to be lying in the range $0.85<H_{p}t_{p}<1.91$, contrary to
the above prediction. This is the so called `age problem' in the
standard flat and inflationary models.

A way out for these models from the age problem is to
postulate that there is a nonzero relic cosmological constant in the
present universe, whose density parameter $\Omega _{\lambda }$ is
comparable to that of matter. But since a cosmological constant
is indistinguishable from the vacuum energy which inflates the
universe, the model is bound to explain how the enormous vacuum
energy, which was present in the early universe gave way to such a
small value at present. This of course will require another extreme
fine tuning and is against the spirit of inflationary models, which
were originally conceived to get rid of all sorts of fine tuning in
the standard model.

This problem is further highlighted in the context of some very
recent measurements \cite{perl} of the deceleration parameter, which
indicates that in the present universe, $q_{p}$ can even be negative.
This can be interpreted as occurring due to the presence of a nonzero
cosmological constant, whose density is comparable to that of the
matter density.  Thus, along with the age problem, the cosmological
constant problem is aggravated by the inflationary models.

Lastly, the singularity problem is not addressed in the inflationary
models. In most models, the inflationary stage is expected to occur
at a time many orders of magnitude greater than the Planck time. Thus
questions like how the universe came into `existence' etc. are not
addressed in the model.

\section {Decaying-$\lambda $ Cosmologies}
\label{sec-decaying}

The cosmological constant problem has triggered a lot of work in the
literature \cite{ozertaha}-\cite{viswa} aimed at a solution based upon
a dynamical $\lambda $; i.e., $\lambda $ or equivalently $\rho
_{\lambda }\equiv \lambda /8\pi G$ varying with time. An
important motivation for considering a variable $\rho _{\lambda }$ can be
explained as follows \cite{nano}: If $\rho _{\lambda }$ corresponds to the
vacuum energy density, then its value is expected to be of the order
of Planck density $\rho_{pl} = 0.5\times 10^{94}$ g cm$^{-3}$ at
least in the Planck epoch. But all observations at present indicate a
very low value $\approx 10^{-29}$ g cm$^{-3}$ for this quantity. An
order of magnitude calculation reveals that since the present age of
the universe is $\approx 10^{61}$ times the Planck time, if $\rho
_{\lambda }$ decays with time from this initial value, then $\rho
_{\lambda ,p} \approx 0.5\times 10^{94} /(10^{61})^{2} \approx
10^{-29}$ g cm$^{-3}$ as expected;  i.e., the cosmological constant
obeys an inverse square law in time. 

All decaying-$\lambda $ models are not precisely of this type.  Here we
discuss two pioneering decaying-$\lambda $ cosmological models
\cite{ozertaha,chen} which propose phenomenological laws for the
time-dependence of $\rho _{\lambda }$. It is of interest to note that
while analysing the thermodynamic correctness of some decaying-$\lambda $
models using the Landau-Lifshitz theory for non-equilibrium
fluctuations, Pavon \cite{pavon} has found that only these two models
successfully pass their test. 
Special cases of these models are obtainable  as
 the new cosmological model to be discussed in the
following chapters of this thesis, which is derived at a more fundamental
level. 

\subsection {Ozer and Taha Model}

In the first of its kind, Ozer and Taha considered a decaying
$-\lambda $ model \cite{ozertaha} in which $\tilde {\rho }=\rho _{m}
+ \rho _{\lambda }$ with $\rho _{m}$ denoting either the relativistic
or nonrelativistic matter. The Einstein equation and the conservation
equation in this case are

\begin{equation}
\frac {\dot {a}^{2}}{a^{2}} + \frac {k}{a^{2}} = \frac {8\pi
G}{3}(\rho _{m}+\rho _{\lambda })
\end{equation}
and

\begin{equation}
\frac {d}{dt} (\rho_{m}a^{3}) + p_{m}\frac {da^{3}}{dt}+a^{3}\frac
{d\rho _{\lambda }}{dt} =0. \label{eq:consot}
\end{equation}

They noted that, for the solution of the cosmological problems, there
should be entropy production and this will require $d\rho _{\lambda
}/dt <0$. Also they argued that, since the present matter density in
the universe is close to the critical density and since these two
are time-dependent terms in the fundamental dynamical equations of
GTR, the equality of $\rho _{m}$ and $\rho _{c}$ would bestow some
special status to the present epoch $t=t_{p}$. Thus they assume that
this equality is a time-independent feature and impose the condition
$\rho _{m} = \rho _{c}$ in the above equations. These conditions
immediately yield

\begin{equation}
k=+1
\end{equation}
and

\begin{equation}
\rho _{\lambda } = \frac {3}{8\pi G} \frac {1}{a^{2}}.
\end{equation}
In the relativistic era in which $p_{m,r}=\frac {1}{3} \rho _{m,r}$, they
obtain a nonsingular solution

\begin{equation}
a^{2}= a_{0}^{2} + t^{2},\label{eq:aot}
\end{equation}
where $a_{0}$ is the minimum value of the
scale factor. The relativistic matter density is

\begin{eqnarray}
 \rho _{m,r} = \rho _{0} \left( \frac {a_{0}}{a}\right) ^{4}-\frac {1}{a^{4}}
\int _{a_{0}}^{a}a^{\prime 4} \frac {d\rho _{\lambda }}{da^{\prime
}}da^{\prime } & = & \frac {3}{8\pi G}\left( \frac {1}{a^{2}}- \frac
{a_{0}^{2}}{a^{4}}\right) \nonumber \\
& = & \frac {3 }{8\pi G}\frac
{t^{2}}{(a_{0}^{2}+t^{2})^{2}}. 
\end{eqnarray}
 
The radiation temperature $T$ is assumed to be related to $\rho _{m,r}$
by the relation (\ref{eq:rhor})

\begin{equation}
\rho _{m,r} = g_{tot}\sigma T^{4},
\end{equation}
from which

\begin{equation}
T= \left( \frac {3 }{8\pi Gg_{tot}\sigma } \right) ^{1/4}\left[ \frac
{t^{2}}{(a_{0}^{2}+t^{2})^{2}} \right] ^{1/4}.
\end{equation}
Thus $T=0$ at $t=0$. The model predicts creation of matter at the
expense of vacuum energy. A maximum temperature $T_{max}$ is attained
at $t= a_{0}$ and is given by

\begin{equation}
T_{max}= \left( \frac {3 c^{4}}{8\pi Gg_{tot}\sigma }\frac
{1}{2a_{0}^{2}} \right) ^{1/4} .
\end{equation}
It was observed that $T_{max}$ should correspond to  the only energy
scale present in the theory, which is the Planck energy and that this
will give $a_{0}\approx 0.03$ $l_{pl}$ (assuming $g_{tot} \approx
100$ in the early relativistic era). Also for $t\gg a_{0}$, the
values of the energy density and temperature attained by radiation at
time $t$ in the standard model are attained at time $2t$ in their
model. Thus it is anticipated that the model has the same thermal
history as the standard model. However, there is difference in the
behaviour of the scale factor and also there is entropy production, which
will help to solve the main cosmological problems. In particular,
they have shown that causality will be established within a time
$t_{caus}\approx 2.3 a_{0}$, soon after the Planck epoch. It was also
shown that the present monopole density in their model is much
smaller than the critical density, which solves  the monopole
problem.

Though solutions are obtained for the pure radiation era with the
above assumptions, they had to impose extra assumptions to determine
the model when nonrelativistic matter is present. It was assumed that
the early pure radiation era soon gave way to a period of matter
generation, where $a_{1}\leq a\leq a_{2}$. After that epoch, i.e.,
when $a\geq a_{2}$, $\rho = \rho _{m,r}+\rho _{m,nr}$ and Eq.
(\ref{eq:consot}) can  be written as

\begin{equation}
d(\rho _{m}a^{3}) + d(\rho _{m,r}a^{3}) + p_{m,r}\; da^{3} = -a^{3}d\rho
_{\lambda }.
\end{equation}
In this era, it is assumed that 

\begin{equation}
\mid \frac {d}{dt}(\rho _{m,nr}a_{0}^{3}\mid \ll \mid \frac {d}{dt}
(\rho _{m,r}a^{3})\mid .
\end{equation}
so that one obtains the solution

\begin{equation}
\rho _{m,r} = \frac {3}{8\pi G}\left( \frac {1}{a^{2}}+\omega \frac
{a_{p}^{2}}{a^{4}}\right),
\end{equation}
where $\omega $ is a dimensionless constant and $a_{p}$ is the
present value of the scale factor. Here, $\rho _{m,nr}a^{3}$, the
total rest mass energy remains a constant. The regions
$a_{0}<a<a_{1}$ where $p_{m,r}=(1/3)\rho _{m,r}$ is separated from
the regions $a\geq a_{2}$ by the epoch of matter creation, which may
be considered as a region of phase transition. The time corresponding
to $a_{1}$ is expected to be $t_{1}\leq 10^{-34}$ s;  i.e., the
GUT era. The reversal of sign of $\ddot {a}$ occurs during this time.

Throughout the evolution, the expression for $p_{\lambda }$ is the
same. Some predictions of the model are independent of the
dimensionless parameter $\omega $. These include $a_{p}\geq
1.578\times 10^{30}$ cm, $\rho _{\lambda }\approx 8.26\times
10^{-30}$ g cm$^{-3}$, $t_{p}\approx (2/3)H_{p}^{-1}$ and
$q_{p}\approx 1/2$. Thus the values of $t_{p}$ and $q_{p}$ are nearly
the same as those of the standard flat model.

It was observed by the authors themselves that the imposition of the
condition $\rho _{m} = \rho _{c}$ is unphysical and that it may be
worthwhile to seek a dynamic principle that determines the form of
$\rho _{\lambda }$, this being the most fundamental assumption made
in the model.

\subsection {Chen and Wu Model}

Chen and Wu \cite {chen}, while introducing their widely discussed
decaying-$\lambda $ cosmological model, have made an interesting argument
in favour of an $a^{-2}$ variation of the effective cosmological
constant on the basis of some dimensional considerations in line with
quantum cosmology. Their reasoning runs as follows: Since there is no
other fundamental energy scale available, one can always write $\rho
_{\lambda}$, the energy density corresponding to the effective
cosmological constant as the Planck density ($\rho _{pl} =
c^{5}/\hbar G^{2} = 5.158\times 10^{93}$ g cm$^{-3}$ ) times a
dimensionless product of quantities.  Assuming that $\rho _{\lambda}
$ varies as a power of the scale factor $a$, the natural ansatz is

\begin{equation}
\rho _{\lambda} \propto \frac {c^{5}}{\hbar G^{2}} \left(
\frac {l_{pl}}{a}\right)^{n} \label{eq:rholchen0}
\end{equation}
One can now  show that $n=2$ is a preferred choice. It is easy to
verify that $n<2$ (or $n>2$) will lead to a negative (positive) power
of $\hbar $ appearing explicitly in the right hand side  of the above
equation. Such an $\hbar $-dependent $\rho _{\lambda }$ would be
quite unnatural in the classical Einstein equation for cosmology much
later than the Planck time.  But it may be noted that $n=2$ is just
right to survive the semiclassical limit $\hbar \rightarrow 0$. This
choice is further substantiated by noting that $n \leq 1$ or $n\geq
3$ would lead to a value of $\rho _{\lambda }$ which violates the
observational bounds. Thus Chen and Wu make the ansatz

\begin{equation}
\rho _{\lambda} = \frac {\gamma }{8\pi G a^{2}}, \label{eq:rholchen} 
\end{equation}
where $\gamma $ is a phenomenological constant parameter.  Assuming
that only the total energy-momentum is conserved, they obtain, for
the relativistic era,

\begin{equation}
\rho _{m,r} = \frac {A_{1}}{a^{4}} + \frac {\gamma }{8\pi G
a^{2}}\label{eq:rhomchen1} 
\end{equation}
and for the nonrelativistic era,

\begin{equation}
\rho _{m,nr} = \frac {A_{2}}{a^{3}} + \frac {2\gamma }{8\pi G
a^{2}}\label{eq:rhomchen2} 
\end{equation}
where $A_{1}$ and $A_{2} $ are to be positive. The Chen-Wu model thus
differs from the standard model in that it has a decaying
cosmological constant and that the matter density has conserving and
nonconserving parts [given by the first and second terms respectively
in equations (\ref{eq:rhomchen1}) and (\ref{eq:rhomchen2})]. By
choosing $\gamma $ appropriately, they hope to arrange $\rho
_{\lambda} $ and the nonconserving parts in $\rho _{m,r}$ and $\rho
_{m,nr}$ to be insignificant in the early universe, so that the
standard model results like nucleosynthesis are undisturbed. But for
the late universe, it can have many positive features like providing
the missing energy density in the flat and inflationary models, etc..
The model predicts creation of matter, but the authors argue that the
creation rate is small enough to be inaccessible to observations.

Conversely to the requirement that the nonconserving parts of matter
density should be negligible in the early universe for standard model
results to remain undisturbed, one can deduce that in this model, the
standard model results are applicable to only the conserving part of
matter density. The nonconserving part is, in fact, created in the
late universe.  Thus for the standard model results to be applicable
to the present universe, the conserving part of the matter density
should be substantial. This in turn will create some problem with
observations.  For example, let us assume that at present, the
conserved part of the nonrelativistic matter density is equal to the
nonconserved part.  Since the vacuum density is only one-half the
nonconserved part [see equations (\ref{eq:rholchen}) and
(\ref{eq:rhomchen2})], for a $k=0$ universe, the deceleration
parameter at present will be $q_{0} = (\Omega _{m}/2)-\Omega
_{\Lambda }= 0.2$.  This is not compatible with the observations
mentioned earlier \cite {perl}.

To avoid the problems in the early universe, they have to assume the
occurrence of inflation, which in turn is driven by the vacuum energy.
But they apply their ansatz to the late-time vacuum energy density
(which corresponds to the cosmological constant) and not to that
during inflation. But the stress energy associated with the vacuum
energy is identical to that of a cosmological constant and it is not
clear how they distinguish them while applying their ansatz.

\chapter{The New Model}
   
 There are a number of instances of the use of complex numbers or
complex analytic functions in GTR \cite{newmann}. Many of these
applications have a common element, namely the analytic continuation
of a real analytic manifold (the spacetime) into the complex,
producing a complex spacetime. One  such complex coordinate
transformation is the Wick rotation of the time-coordinate $t$ in
Minkowski metric to obtain a Euclidean metric. Similarly the
transformation of one solution of Einstein equation into another by
means of a complex switch of coordinates is well known. For example,
open and closed Friedmann models, de Sitter and anti de Sitter
spacetimes, Kerr and Schwarschild metrics etc. are related by complex
substitutions \cite{fla}. The use of complex variables extends to
more sophisticated ones like spin-coefficient formalism, Ashtekar
formalism, Twistor theory etc. We present our model based upon the
signature change of the metric from Lorentzian (+ - - -) to Euclidean
(++++). A signature change in the early universe is a widely
discuused idea in current literature.  After briefly reviewing the
same, we present our model, which has a direct bearing on many
cosmological observations, a feature unparalleled in most other
applications mentioned above. We discuss the physical model including
its predictions and then show how the model is devoid of cosmological
problems.

\section {Signature Change}
\label{sec-signature}

The Hartle-Hawking `no boundary' condition \cite{hh} in quantum
cosmology allows a change of signature in the Planck epoch, resulting
in the origin of the universe in a regime where there is no time.
(The spacetime metric is Euclidean, so that spacetime is purely
spatial).  Ellis {\sl et al.} \cite{ellis} investigated such a
possibility in the classical solution of the Einstein field
equations. They argue that the usual solutions of this equation with
Lorentzian metric are
not because it is demanded by the field equations, but rather because
it is a condition we impose on the metric before we start looking for
solutions. They obtain a classical signature change by replacing the
squared lapse function $N^{2}(t)$ appearing in the metric in the ADM
3+1 split of RW spacetime (See Secs. \ref{sec-gtr},
\ref{sec-homoiso}) with $\nu$  and allowing it to be negative.
During signature change, $\nu $ passes through the value zero (which
is a form of singularity) and hence a crucial point is the matching
conditions at the surface of change. Ellis {\sl et al.} have claimed
to obtain RW solutions of the classical Einstein equations where $\nu
$ changes sign at some time $t_{0}$, with the condition that the
matter density and pressure are finite and the 3-space metric $h_{\mu
\nu }$ is regular, as the change of sign takes place. This condition
is equivalent to requiring that the extrinsic curvature $K_{\mu \nu}$
is continuous at the surface of change. In another approach, Hayward
\cite{hayward1} obtained signature changing solutions by requiring
that at the surface of change, $K_{\mu \nu }$ should vanish. The
issue of these `junction conditions' is a matter of hot debate in the
current literature.

Another  development in connection with the signature of the metric
was that made by Greensite \cite{green}, who proposed a dynamical
origin for the Lorentzian signature. The idea is to generalize the
concept of Wick rotation in path integral quantisation. Rather than
viewing Wick rotation as a mathematical technique for the convergence
of the path integral, the Wick angle $\theta $ is treated as
dynamical degree of freedom. He claims to have obtained a relation
between the dimension and signature of spacetime, which favour a
Lorentzian signature for a 4-dimensional spacetime and explain the
presence of the factor $i$ in the path integral amplitude. As a more
general approach to signature change, Hayward \cite{hayward2}
extended the idea of Greensite and allowed the lapse function to be
complex. This is claimed to yield a complex action that generates
both the usual Lorentzian theory and its Riemannian analogue and
allows a change of signature between the two.

\medskip 
\section {Derivation of the New model}
\label{sec-new}

We obtain a signature changing RW solution by a different route than
those mentioned above \cite{jj1,jj2}.  If we make a substitution $a(t)
\rightarrow \hat{a}(t) = a(t)e^{i\beta }$ in  Eq. (\ref{eq:rwle}),
then the spacetime has Lorentzian signature (+ - - -) when $\beta  =
\pm n \pi$,  $ ( n = 0,  1,  2, ..)$, and  has Riemannian signature
(++++) when $\beta  =\pm  (2n +1)\pi /2$, $( n = 0,  1,  2, ..)$.
Let  the  solution $a(t)$  be   in   the     form $a_{o} e^{\alpha
(t)}$ . Then the above expression becomes $\hat{a}(t) = a_{o}
e^{\alpha (t)+i\beta }$.  We note that interesting physics appears if
 the time-dependence of the scale factor is shared also by $\beta
$; i.e., $\beta  = \beta (t)$, an assumption consistent with the
homogeneity and isotropy conditions.  Then the signature of the
metric changes  when $ \beta $ varies from $0 \rightarrow \pi /2$
etc.  Our  ansatz   is   to replace $a(t)$ in metric (\ref{eq:rwle})
with 

\begin{equation}
\hat{a}(t) = a(t) e^{i\beta (t)} = a_{o} e^{\alpha (t)+
i\beta (t)} \equiv  x(t) +i\; y(t). \label{eq:ac1}
\end{equation}
We further assume that this model of the universe with a complex scale factor 
is closed (i.e.,$ k = +1$) and has a zero energy-momentum tensor (i.e.,
$I_{M}=0$).
Thus we start with  a system obeying an action principle,
where the action is given by
\par
\begin{equation}
I = \frac{-1}{16\pi G}\int  (-g)^{1/2} { R} \; d^{4}x. \label{eq:complexi}
\end{equation}
Here 

\begin{equation}
{R} = \frac {6}{N^{2}}\left( \frac {\dot{\hat{a}}}{\hat{a}}\right) ^{2} 
-\frac {6}{\hat{a}^{2}} ,
\end{equation}
 an expression similar to (\ref{eq:ricci}).
 Using this and integrating the space part, we get Eq.
(\ref{eq:complexi}) as 
 
 \begin{equation}
I = - \frac {3\pi}{4G} \int N \; {\hat {a}^{3}} \left[ 
\frac {1}{N^{2}} \left( \frac { \dot {\hat {a}}}{\hat {a}}\right)^{2} - \frac
{1}{{\hat {a}^{2}}} \right] dt \label{eq:cact}
 \end{equation}
 Minimising this  action with respect to variations 
 of $N$ and $\hat {a}$  and fixing the gauge 
$ N = 1 $, we get the constraint and the field equations

\begin{equation}
\left( \frac { \dot {\hat {a}}}{\hat {a}} \right) ^{2} + 
\frac {1}{\hat {a}^{2}} = 0 \label{eq:t-tc}
\end{equation}
and

\begin{equation}
2\frac {\ddot {\hat {a}}}{\hat {a}} + \left( 
\frac {\dot {\hat {a}}}{\hat {a}}\right) ^{2} + \frac {1}{\hat
{a}^{2}} = 0. \label{eq:s-sc} 
\end{equation}
respectively. With $\hat {a}(t) \equiv x(t) +i\; y(t) $ and $ x_{0}$,
$ y_{0} $ constants, these equations may be solved to get

\begin{equation} 
\hat {a}(t) = x_{0} + i\; (y_{0} \pm t).
\end{equation}
We can choose the origin $ t= 0 $ such that $\hat {a} (0) = x_{0}$.
Relabelling $x_{0} \equiv a_{0}$, we get,
\par
\begin{equation}
\hat {a} (t) = a_{0} \pm i\; t . \label{eq:ac2}
\end{equation}
This equation gives the contour of evolution of $\hat {a}(t)$ which
is a straight line parallel to the imaginary axis. At $t=0$, this
leaves the signature of spacetime Lorentzian but as $t \rightarrow
\infty $ it becomes almost Riemannian. This need not create any
conceptual problem since here we are considering only an unperceived
universe with zero energy-momentum tensor whose existence is our
ansatz.  (Simple physical intuition would give a signature
`Riemannian at early times and Lorentzian at late' if it was for the
physical universe we live in with matter contained in it. But in the
above, we have a signature change in the opposite manner for the
unphysical universe devoid of matter and this need not contradict our
physical intuition).  The connection with a closed real or physical
universe is obtained by noting from the above that
\par
\begin{equation}
a^{2}(t) = \mid \hat {a} (t) \mid ^{2} = a^{2}_{0} + t^{2}.
\label{eq:aotc} 
\end{equation}
This is the same equation (\ref{eq:aot}) which governs the evolution
of scale factor in the relativistic era of the Ozer-Taha model
\cite{ozertaha}. But in that model, $a_{0}$ is undetermined; as
mentioned in Sec. \ref{sec-decaying}, it is only speculated to be of
the order of Planck length. In our case, a quantum cosmological
treatment to follow in Sec. \ref{sec-complexq} reveals that
$a_{0}=\sqrt {2G/3\pi }\approx l_{pl}$.

\section {The Real Universe}
\label{sec-real} 

 Separating the real and imaginary parts of  (\ref{eq:t-tc}) and
(\ref{eq:s-sc}) and combining them, one easily obtains the following
relations:

\begin{equation}
\frac {\dot {a}^{2}}{a^{2}} + \frac {1}{a^{2}} = \dot {\beta }^{2} +
\frac {2}{a^{2}} \sin ^{2}\beta , \label{eq:t-tbeta}
\end{equation}

\begin{equation}
2\frac {\ddot {a}}{a}+\frac {\dot {a}^{2}}{a^{2}} + \frac {1}{a^{2}}
= 3\left( \dot {\beta }^{2} + 
\frac {2}{3a^{2}} \sin ^{2}\beta \right) , \label{eq:s-sbeta}
\end{equation}

\begin{equation}
\ddot {\beta } + 2\dot {\beta }\frac {\dot {a }}{a} =0,
\label{eq:consbeta} 
\end{equation}
and

\begin{equation}
2\dot {\beta }\frac {\dot {a}}{a} = \frac {1}{a^{2}}\sin 2\beta
.\label{eq:betadot} 
\end{equation}
Also from (\ref{eq:ac1}) and (\ref{eq:ac2}), we get

\begin{equation}
\beta (t) = \tan ^{-1} ( \frac {\pm t} {a_{0}})
\end{equation}
and 

\begin{equation}
\dot {\beta }(t) = \frac {\pm a_{0}}{a^{2}(t)} = 
\frac {\pm \cos ^{2} \beta }{a_{0}}.
\end{equation}
With the help of the last two equations we observe that the real parts of 
 (\ref{eq:t-tc}) and  (\ref{eq:s-sc}) can be rewritten in terms 
of $a= a(t) = \mid \hat {a}(t) \mid $ as

\begin{equation}
\frac  {\dot {a}^{2}}{a^{2}}  + \frac {1}{a ^{2}} = \frac {2}{a^{2}} - \frac 
{a^{2}_{0}}{a^{4}}  \label{eq:constrainte}
\end{equation}

\begin{equation}
2\frac {\ddot {a}}{a}+\frac {\dot {a}^{2}}{a^{2}}  + \frac {1}{a ^{2}} =
\frac {2}{a^{2}} + \frac  
{a^{2}_{0}}{a^{4}}, \label{eq:fe}
\end{equation}
whose solution is the same as that obtained in
(\ref{eq:aotc}).  
We see that the real quantity $a(t)$ may be considered as the scale
factor of a nonempty RW universe. Eqs. (\ref{eq:constrainte}) and
(\ref{eq:fe}) are appropriate 
for a closed RW model with real scale factor $a$ and with total
energy density and total pressure  given by [See Eqs. (\ref{eq:t-t})
and (\ref{eq:s-s})], 

\begin{equation}
\tilde{\rho}  = \frac{3}{8\pi G}\left( \frac{2}{a^{2}} -
\frac{ a^{2}_{0}}{a^{4}}\right) ,\label{eq:rhot}
\end{equation}

\begin{equation}
\tilde{p}   ={-\frac{1}{8\pi G}}\left( \frac {2}{a^{2}} + 
\frac{a^{2}_{0}}{a^{4}} \right) \label{eq:pt}
\end{equation}
respectively, whose breakup can be performed in many ways. 

\subsubsection {Matter and Vacuum}

First let us assume, as done in \cite{ozertaha}, that

\par
\begin{equation}
\tilde{\rho}  =  {\rho} _{m} +  {\rho} _{\lambda }, \label{eq:rhotml}
\end{equation}
\begin{equation}
\tilde{p} =  {p}_{m} +  {p}_{\lambda }. \label{eq:ptml}
\end{equation}
  We     write the equations of state in the form
\par
\begin{equation}
 {p}_{m} = w\;  {\rho} _{m} \label{eq:esm}
\end{equation}
and

\begin{equation}
 {p}_{\lambda } = -  {\rho} _{\lambda }. \label{eq:esl}
\end{equation}
 Solving (\ref{eq:rhot})  and (\ref{eq:pt}) using
(\ref{eq:rhotml})-(\ref{eq:esl}),  one gets
\par
\begin{equation}
 {\rho} _{m} = \frac{4}{8\pi G(1+w)}\left( \frac{1}{a^{2}} -
 \frac{a^{2}_{o}}{a^{4}}\right) , \label{eq:rhom1}
\end{equation}
\medskip
\begin{equation}
 {\rho _{\lambda }}  = \frac{1}{8\pi G(1+w)}\left[ \frac {2(1+3w)}{a^{2}} + 
 \frac {a^{2}_{0}(1-3w)}{a^{4}}\right] . \label{eq:rhol1}
\end{equation}
For  a relativistic matter dominated universe,  the  matter  density 
 and the vacuum density  are
\par
\begin{equation}
 {\rho }_{m,r}  = \frac {3}{8\pi G}\left( \frac {1}{a^{2}} - 
 \frac{a^{2}_{0}}{a^{4}}\right) ,\label{eq:rhomr1}
\end{equation}

\begin{equation}
 {\rho} _{\lambda ,r}  =\frac {3}{8\pi G}\frac {1}{a^{2}}. \label{eq:rholr1}
\end{equation}
From (\ref{eq:fe}),  the critical density of the  real universe is
\par
\begin{equation}
\rho_{c}  \equiv \frac{3}{8\pi G} H^{2}  = \frac{3}{8\pi G}
\left( \frac{1}{a^{2}} - 
\frac{a^{2}_{o}}{a^{4}}\right) , 
\end{equation}
where $ H $ , the value of the Hubble parameter 
is assumed to coincide with that predicted by the model. (We can see  that 
this is indeed the case in the present epoch by evaluating the combination 
$H_{p}t_{p}\equiv \left[ \frac {\dot {a}} {a}\right] _{p} t_{p}$ 
for $a_{p} \gg a_{0}$, which is found to be nearly equal to unity.) 
Then the ratios of  density  to  critical  density  for  matter  and
vacuum energy in the relativistic era are
\par
\begin{equation}
 {\Omega} _{m,r}\equiv \frac { {\rho }_{m,r}}{ {\rho }_{c}}=1, \label{eq:omr1}
\end{equation}

\begin{equation}
 {\Omega}_{\lambda ,r}\equiv \frac{ {\rho}_{\lambda ,r}}{ {\rho}_C}
 \approx 1\qquad \hbox {for} \qquad a(t) \gg a_{0}. \label{eq:olr1}
\end{equation}
For a universe dominated by nonrelativistic  matter,  the  condition
$w = 0$ may  be used in (\ref{eq:rhom1}) and (\ref{eq:rhol1}). In
this case,
\par
\begin{equation}
 {\Omega}_{m,nr}=4/3,
\end{equation}

\begin{equation}
 {\Omega} _{\lambda ,nr} \approx  2/3 \qquad  \hbox {for }
 \qquad a(t) \gg a_{0}
\end{equation}
It  may be noted that (\ref{eq:rhomr1}) and (\ref{eq:rholr1}) are the
same  expressions as those  obtained  in \cite{ozertaha} and
(\ref{eq:omr1}) is their ansatz.  But  the last two results  for  the
nonrelativistic era are outside the scope of that model.
\par

\subsubsection {Matter, Vacuum and Negative Energy}

In the above, we have assumed $\tilde {\rho } = \rho _{m} + \rho
_{\lambda } $ following the example in \cite{ozertaha}. But this
splitup is in no way unique. Equation (\ref{eq:rhom1}) gives $\rho
_{m} = 0$     at $t$ = 0.  In order to   avoid this less probable
result, we  assume that the term $-(3/8\pi G) (a_{0}^{2}/a^{4})  $ in
$\tilde{\rho} $ is  an  energy  density appropriate   for   negative
energy relativistic  particles. The  pressure   $ {p}\_$
corresponding to this negative energy density $ {\rho } \_ $  is also
negative.  Negative energy densities in the universe were postulated
earlier \cite{hn}.  Such an assumption has the further advantage of
making the expressions for $\rho _{m} $ and $\rho _{\lambda }$ far
more simple and of conforming  to the Chen  and Wu \cite{chen}
prescription of  a pure $a^{-2}$ variation of vacuum density
(though the Chen-Wu arguments, with $a_{0}$
identified as the Planck length, are not against the form
(\ref{eq:rhol1}) for $\rho _{\lambda }$ since the term which contains
$a_{0}^{2}/a^{4}$ becomes negligibly small when compared to the
$a^{-2}$ contribution within a few Planck times).   Thus we use a
modified ansatz in this regard [instead of (\ref{eq:rhotml})-
(\ref{eq:esl})],
\par
\begin{equation}
\tilde{\rho}  =  {\rho} _{m} +  {\rho} _{\lambda } +  {\rho} \_ ,
\label{eq:rhotml1} 
\end{equation}
\begin{equation}
\tilde{p} =  {p}_{m} + { p}_{\lambda } +  {p}\_ ,
\end{equation}
\begin{equation}
{p}_{m} = w\;  {\rho} _{m},
\end{equation}
\begin{equation}
 {p}_{\lambda } = -  {\rho} _{\lambda },
\end{equation}
\begin{equation}
 {p}\_ = \frac{1}{3}  {\rho} \_ ,
\end{equation}
and
\par
\begin{equation}
 {\rho} \_ = - \frac{3}{8\pi G}\frac{a^{2}_{o}}{a^{4}} \label{eq:rho-}
\end{equation}
and solving  (\ref{eq:rhot}) and  (\ref{eq:pt})  with   these
choices,   the results are,
\par
\begin{equation}
 {\rho}_{m}  =\frac{4}{8\pi G(1+w)}\frac{1}{a^{2}}, \label{eq:rhom2}
\end{equation}
\begin{equation}
 {\rho}_{\lambda }  =\frac{2(1+3w)}{8\pi G(1+w)}\frac{1}{a^{2}}
\label{eq:rhol2} 
\end{equation}
so that
\par
$$
 {\Omega} _{m,r} \approx  1 \qquad  {\Omega} _{\lambda ,r} \approx  1,
$$
\begin{equation}
 {\Omega} _{m,nr} \approx 4/3 \qquad  
 {\Omega} _{\lambda ,nr} \approx  2/3
\end{equation}
$$
 {\Omega}\_\equiv\frac { {\rho}\_}{\rho_{C}} \ll 1 
$$
for $a(t) \gg a_{0} $. The predictions for $\Omega_{m}$ are marginal, though 
not ruled out by observations. 

\subsubsection {Matter, Vacuum Energy, Negative Energy and K-matter}

Many authors \cite{kolb,kam,string}
seriously consider the existence 
of a new form of matter in the universe (called K-matter \cite{kolb} - 
perhaps a stable texture \cite{kam}) 
with the equation of state $ p_{K} = - \frac {1}{3} \rho _{K} $ and which 
decreases  as $a^{-2}$.  This leads to the idea of a low density closed 
universe \cite{kam}. If we accept this as probable,  the prediction 
 for $ \Omega_{m} $ will be well within 
the observed range of values. In this case we include a term 
$ \frac {3}{8\pi G}
\frac {K}{a^{2}}$ to the right hand side of (\ref{eq:rhotml1}) so that

\begin{equation}
\rho _{m} = \frac {2}{8\pi G} \frac {1}{(1+w)} \frac
{(2-K)}{a^{2}}\label{eq:rhom3} 
\end{equation}
and 

\begin{equation}
\rho _{\lambda } = \frac {1}{8\pi G } \frac {(1+3w)}{(1+w)} 
\frac {(2-K)}{a^{2}}\label{eq:rhol3}
\end{equation}
For a typical value $ K=1$ \cite{kolb}, the predictions for $a\gg a_{0}$ are

$$
\Omega _{m,r} \approx 1/2, \qquad \Omega _{\lambda , r} \approx 1/2,
$$

\begin{equation}
 \Omega _{m, nr} \approx 2/3, \qquad \Omega _{\lambda . nr}
\approx 1/3, 
\end{equation}

$$
 \Omega \_ \ll 1, \qquad \Omega _{K} \approx 1
$$
\medskip

The model makes clear cut predictions regarding the total energy
density $\tilde {\rho}$ and total pressure $\tilde {p}$ as given by
(\ref{eq:rhot}) and (\ref{eq:pt}) but the decomposition of these do
not follow from any fundamental principles, except for those
heuristic reasons we put forward.  It is easy to see that the
conservation law for total energy
\par
\begin{equation}
\frac{d(a^{3}\tilde{\rho})}{dt} = -\tilde{p}\;\frac{da^{3}}{dt}
\end{equation}
is  obeyed, irrespective of  the  ansatze   regarding   the
detailed structure of $\tilde{\rho} $.
\par
\medskip

\section {Thermal Evolution and Solution of Cosmological Problems}
\label{sec-thermal}

One can see that the solution of cosmological problems mentioned in
Sec. \ref{sec-problems} does not
significantly depend on the split up of $\tilde {\rho}$. Note that in
all the above cases, $\Omega _{m}$ is time-independent  when
$a\gg a_{0}$. Not only $\Omega _{m}$, but also all the density
parameters including the total density parameter (which may be
defined as $\tilde {\Omega }\equiv \tilde {\rho }/\rho _{c}$) are
constants in time. This is not difficult to understand: for $a\gg
a_{0}$, $\tilde {\rho }$ and all other densities vary as $a^{-2}$.
This, when put in (\ref{eq:flat}) tells us that $\tilde {\Omega }$ and
 all other density parameters are time-independent for large
$t$. Thus there is no flatness problem in this model.

Another notable feature is that in all the above cases, we have $\rho
_{m}/\rho _{\lambda } = 2$ in the nonrelativistic era.
 Thus the  model predicts that the energy density 
corresponding to the cosmological constant is comparable with matter
density and this solves the cosmological constant problem too. It can
also be seen that according to the model, the observed universe
characterised by the present Hubble radius has a size equal to the
Planck length at the  Planck epoch and this indicates that the
problem with the size of the universe does not appear here. 

Next let us consider the horizon problem. A necessary condition for
the solution of this problem before some time $t_{s}$ is given by Eq.
(\ref{eq:hcond}). Using our expression (\ref{eq:aotc}) for the scale
factor with $a_{0} \approx l_{pl}$, we see that the horizon problem
is solved immediately after the Planck epoch, even if we extend the
lower limit of integration in (\ref{eq:hor}) to $t_{pl}$.  For the
investigation of other problems, we have to study the thermal
evolution of the universe as envisaged in the model.

The relativistic matter density in the present model [using
(\ref{eq:rhom2}) or more generally (\ref{eq:rhom3})] can be written
as

\begin{equation}
\rho _{m,r} = \frac {3}{8\pi G} \frac {\kappa }{a^{2}} = \frac {3}{8\pi
G} \frac {\kappa }{a_{0}^{2}+t^{2}} , \label{eq:rhom4}
\end{equation}
where $\kappa = 1 - \frac {K}{2} $ is a constant of the order of unity.
Using (\ref{eq:rhor}) we find the corresponding temperature as 

\begin{equation}
T = \left( \frac {3}{8\pi G} \frac {\kappa } 
{g_{tot}\sigma } \right) ^{1/4} 
\left( \frac {1} {a_{0}^{2} + t^{2}} \right) ^{1/4}, \label{eq:t4}
\end{equation}
which is a maximum at $t=0$. (In natural units $\sigma =
\pi^{2}/30$.) If $a_{0} =\sqrt {2G/3\pi }$ as mentioned in
Sec. 3.2, then $ T(0) \approx 0.36 \times \kappa ^{1/4}G^{-1/2} $, which
is comparable with the Planck energy and as $t \rightarrow \infty $,
T decreases monotonically.

The above expressions (\ref{eq:rhom4}) and (\ref{eq:t4}) may be
compared with the corresponding expressions in the standard model:

\begin{equation}
\rho _{s.m.} = \frac {3}{8\pi G} \frac {1}{ (2t)^{2}},
\end{equation}

\begin{equation}
T_{s.m.}=\left[ \frac {3}{8\pi G}\frac {1 }{g_{tot}\sigma }.
\right]^{1/4} \frac {1}{(2 t)^{1/2}}
\end{equation}
Assuming that $\kappa ^{1/4}$ is close to unity,
it can be inferred that the values of $\rho _{m,r}$ and $T$ attained at
time $t$ in the standard model are attained at time $\sqrt {2} t$ in
the present model. Thus the thermal history in the present model is
expected to be essentially the same as that in the standard model.
But the time-dependence of the scale factor is different in our
model; we have a nearly coasting evolution 
and this helps us to solve the cosmological problems.

It can now be shown that density perturbations on scales well above the
present Hubble radius can be generated in this model  by
evaluating the communication distance light can travel between
the Planck time $t_{pl}$ and $t_{dec}$, the time of decoupling
\cite{liddle}: 

\begin{equation}
d_{comm} (t_{pl}, t_{dec}) = a_{p} \int _{t_{pl}}^{t_{dec}} \frac
{dt}{a(t)} = 0.62 \times 10^{6} {\hbox {Mpc}}
\end{equation}
where we have used $t_{dec} \approx 10^{13}s$,  the same as
that in the standard model, an assumption which is justifiable on the
basis of our reasoning made before regarding thermal history.  Thus
the  evolution in our case has the communication distance between
$t_{pl}$ and $t_{dec}$ much larger than the present Hubble radius and
hence it can generate density perturbations on scales of that order.
[See Eq. (\ref{eq:comm2}).] It is interestng to note that Liddle
\cite {liddle} has precluded coasting evolution as a viable means to
produce such perturbations and argued that only inflation ($\ddot {a}
>0$) can perform this task, thus "closing the loopholes" in the
arguments of Hu {\sl et al.} \cite {hu}. But it is worthwhile to
point out that his observations are true only in a model which coasts
from $t_{pl}$ to $t_{nuc}$   and thereafter evolves according to the 
standard model (See Sec. \ref{sec-attempts}). In our case, the
evolution is coasting throughout the 
history of the universe (except during the Planck epoch) and hence
his objection is not valid.

A bonus point of the present approach, when compared to standard and
inflationary models may now be noted. In these  models, the
communication distance between $t_{nuc}$ and $t_{dec}$, or for that
matter the communication distance from any time after the production
of particles (assuming this to occur at the end of inflation) to the
time $t_{dec}$ will be only around $200 h^{-1}$ Mpc \cite{liddle}.
Thus density perturbations on scales above the Hubble radius cannot
be generated in these models in the period when matter is present.
This is because inflation cannot enhance the communication distance
after it. The only means to generate the observed density
perturbations is then to resort to quantum fluctuations of the
inflaton field. The present model is in a more advantageous position
than the inflationary models in this regard since the communication
distance between $t_{nuc}$ and $t_{dec}$ in this case is

\begin{equation}
d_{comm} (t_{nuc}, t_{dec}) = a_{p} \int _{t_{nuc}}^{t_{dec}} \frac
{dt}{a(t)} = 4.35\times 10^{29} \hbox {cm} =1.45 \times 10^{5} \hbox {Mpc}
\end{equation}
which is again much greater than the present Hubble radius. So we can
consider the generation of the observed density perturbations as a
late-time classical behaviour too.

It can be seen that entropy is produced at the rate 

\begin{equation}
\frac {dS}{dt} = 4\pi ^{2} \frac {3 \kappa } {8\pi G} \left[  \frac
{8\pi G } {3}  \frac {g_{tot}\sigma }{\kappa } \right] ^{1/4} 
\frac {t}{(a_{0}^{2} + t^{2})^{1/4}}
\end{equation}
which enables the solution of cosmological problems.
  
Lastly, the present monopole density predicted in this case can be seen to 
be [See Eq. (\ref{eq:mono})]

\begin{equation}
n_{monopole} (t_{P}) \approx \frac {3}{4\pi } a^{-3}(t_{p}) \approx
\frac {3}{4\pi }\times 10^{-84} {\hbox {cm}}^{-3}
\end{equation}
so that 

\begin{equation}
\rho _{monopole} \approx \frac {3 }{2\pi } \times 10^{-92} \hbox {g cm}^{-3}
\end{equation}
 This is very close to that estimated in \cite{ozertaha}, 
and is negligibly smaller than the critical density. Thus the monopole
problem is solved also in this case.

\medskip
Irrespective of the case we are considering, the model is nonsingular
and there is no singularity problem. The solution of the age problem
is also generic to the model. It may be noted that the model
correctly predicts the value of the combination $H_{p} t_{p} \approx
1 $.  This places the present theory in a more advantageous position
than the standard flat and the inflationary models with a zero
cosmological constant, where this value is predicted to be equal to
2/3, which is not in the range of recently observed values.

 Another  interesting feature  is  that  since   the   expansion
process is reversible and the basic equations   are  time   reversal
invariant,  we can extrapolate to $t <0$. This  yields   an   earlier
contracting phase  for the universe. Such a phase  was  proposed  by
Lifshitz and Khalatnikov \cite{lif}. If there was such  an  initial
phase, causality  could  have  established itself much earlier   than
the time   predicted in \cite{ozertaha}.
\par
The   model predicts creation  of  matter  at   present with a rate
of creation per unit volume given by
\par
\begin{equation}
\frac{1}{a^{3}}\frac{d(a^{3} {\rho} _{m})}{dt} \mid_{p}\; =  
{\rho} _{m,p}\;H _{p},
\end{equation}
where $ {\rho} _{m,p}$  is the present matter density.  In arriving
at this  result,   we   have made  use of the assumption of a
nonrelativistic  matter   dominated universe. Note that the creation
rate is only  one third of that in the steady-state cosmology
\cite{narlikar}.  Since   the possibility of creation of matter or
radiation at  the required rate cannot be ruled out at the present
level of observation \cite{chen}, this does not pose any serious
objection.

\section{Alternative Approach}     
\label{sec-alter}

We present an alternative  model to the above without resorting to
any complex metric, while preserving all the positive features of the
physical universe envisaged in it, except the avoidance of
singularity. We do this by  modifying the Chen-Wu argument (See Sec.
\ref{sec-decaying}) to include the conserved total energy density
$\tilde {\rho}$ of the universe in place of the vacuum density and
this again brings in some fundamental issues which  need serious
consideration. If the Chen-Wu ansatz is true for $\rho _{\lambda }$,
then it should be true for $\tilde {\rho }$ too. In fact, this ansatz
is better suited for $\tilde {\rho}$ rather than $\rho _{\lambda }$
since the Planck era is characterised by the Planck density for the
universe, above which quantum gravity effects become important. Hence
one can generalise  (\ref{eq:rholchen0}) to write

\begin{equation}
\tilde{\rho } =A \frac {c^{5}}{\hbar G^{2}}\left( \frac
{l_{pl}}{a}\right) ^{n} \label{eq:rhotchen1}
\end{equation}
where $A$ is a dimensionless proportionality constant.  When $\tilde
{\rho}$ is the 
sum of various components and each component is assumed to vary as a
power of the scale factor $a$, then the Chen-Wu argument can be
applied  to conclude that $n=2$ is a preferred choice for each
component. Violating this will force the inclusion of $\hbar $
-dependent terms in $\tilde {\rho }$, which would look unnatural in a
classical theory.  Not only for the Chen and Wu model, in all of
Friedmann cosmologies, this argument may be used to forbid the
inclusion of substantial energy densities which do not vary as
$a^{-2}$ in the classical epoch.

At first sight, this may appear as a grave negative result. But
encouraged by our results in the previous sections, we proceed to the
next logical step of investigating the implications of an $a^{-2}$
variation of $\tilde {\rho}$.  If the total pressure in the universe
is denoted as $\tilde {p}$,  the above result that the conserved
quantity $\tilde {\rho }$ in the Friedmann model varies as $a^{-2}$
implies $\tilde {\rho } + 3 \tilde {p} =0$. This will lead to a
coasting cosmology. Components with such an equation of state are
known to be strings or textures \cite{kam}. Though such models are
considered in the literature, it would be unrealistic to consider our
present universe as string dominated. A crucial observation which
makes our model with $ \tilde {\rho }$ varying as $ a^{-2}$ realistic
is that this variation leads to string domination only if we assume
$\tilde {\rho }$ to be unicomponent. Instead, if we assume that
$\tilde {\rho }$ consists of parts corresponding to relativistic/
nonrelativistic matter and a time-varying cosmological constant,
i.e., if we assume

\begin{equation}
\tilde {\rho } = \rho _{m} + \rho _{\lambda }, \qquad \tilde {p} =
p_{m} + p_{\lambda }\label{eq:rhop},
\end{equation}
then the condition $\tilde {\rho } + 3\tilde {p} =0$ will give

\begin{equation}
\frac {\rho _{m}}{\rho _{\lambda }} = \frac {2}{1+3w}\label{eq:rhom/l}
\end{equation}
In other words, the modified Chen-Wu ansatz leads to the conclusion
that if the universe contains matter and vacuum energies, then vacuum
energy density should be comparable to matter density.  This, of
course, will again lead to a coasting cosmology, but  a
realistic one when compared to a string dominated universe.

$\rho _{m}$ or $\rho _{\lambda }$, which varies as $a^{-2}$, may
sometimes be mistaken for  strings but it should be noted that the
equations of state we assumed for these quantities are different from
that for strings and are what they ought to be to correspond to
matter density and vacuum energy density respectively.  It is true
that components with equations of state $p=w\; \rho $  should obey $
\rho \propto a^{-3(1+w)}$, but this is valid when those components
are separately conserved. In our case, we have only assumed that the
total energy density is conserved and not the parts corresponding to
$\rho _{m}$ and $\rho _{\lambda } $ separately.  Hence there can be
creation of matter from vacuum, but  again the present creation rate
is too small to make any observational consequences.

The solution to the Einstein equations in a Friedmann model with
$\tilde {\rho } + 3\tilde {p}=0$, for all the three cases $k=0, \pm
1$, is the coasting evolution

\begin{equation}
a(t) = \pm m t\label{eq:achen}
\end{equation}
where $m$ is some proportionality constant. The total energy density is
then 

\begin{equation}
\tilde {\rho } = \frac {3}{ 8\pi G} \frac {(m^{2} +k)}{a^{2}}.
\label{eq:rhotchen2} 
\end{equation}
Comparing this with (\ref{eq:rhotchen1}) 
(with n=2), we get $m^{2} +k = 8\pi A/3$.

The prediction regarding the age of the universe in the model is
obvious from Eq. (\ref{eq:achen}). Irrespective of the value of $m$,
we get the combination $H_{p}t_{p}$ as equal to unity, which is well
within the bounds. Thus there is no age problem in this model. We can
legitimately define the critical density as $\rho _{c} \equiv (3/8\pi
G) (\dot {a}^{2}/a^{2})$, so that equation (\ref{eq:rhotchen2}) gives

\begin{equation}
\tilde {\Omega } \equiv \frac {\tilde {\rho }}{\rho _{c}} = \left( 1- \frac
{3k}{8\pi A }\right) ^{-1}
\end{equation}
As in the standard model, we have $\tilde {\Omega } = 1$ for $k=0$
and $\tilde {\Omega }>1 $ ($<1$) for $k= +1$ ($k=-1$).  But unlike
the standard model, $\tilde {\Omega }$ is a constant. Also for $A$
greater than or approximately equal to 1,  we have $\tilde {\Omega
}$ close to unity for all values of $k$.  Using equation (\ref{eq:rhop})
and (\ref{eq:rhom/l}), we get

\begin{equation}
\Omega _{m} \equiv \frac {\rho _{m}}{\rho _{c}} = \frac {2\tilde
{\Omega }}{3(1+w)}, \qquad \Omega _{\lambda } \equiv \frac {\rho
_{\lambda }}{\rho _{c}} = \frac {(1+3w) \tilde {\Omega }}{3(1+w)}
\end{equation}

It is clear that we regain our model in the previous section when
$m=1$ and $k=+1$. In that special case, $A=3/4\pi $ and the present
alternative model is precisely the same as the former, except for the
initial singularity and the evolution in the Planck epoch. But even
when $A$ is not exactly equal to $3/4\pi $ and is only of the order
of unity, the thermal evolution is almost identical and the absence
of cosmological problems is generic to these models.

\chapter{Quantum Cosmology}

Among the fundamental interactions of nature, gravity stands alone;
it is linked to geometry of spacetime by GTR while the other
interactions are describable by quantum fields which propagate in a
`background spacetime'. Another reason is that whereas quantum field
theory assumes a preferred time coordinate and a previlaged class of
observers, GTR demands equivalence among all coordinate systems. Also
in quantum theory,  there is the issue of `observation': the quantum
system is supposed to interact with an external observer who is
described by classical physics, but such notions are alien to GTR. To
sum up, we can say that till now these two major physical theories
remain disunited. Quantum gravity is an attempt to reconcile them. It
is not yet clear what a quantum theory of gravity is, and  there are
several directions pursued in this regard. Perhaps the simplest
application of quantum gravity is in cosmology. The most well studied
approach in quantum cosmology is the canonical quantisation  in which
one writes a wave equation for the universe, analogous to the
Schrodinger equation. This procedure requires a Hamiltonian
formulation of GTR. In this chapter, we present a brief review of
quantum cosmology and then apply the formalism to the cosmological
models discussed in the last chapter.

\section{Hamiltonian Formulation of GTR}
\label{sec-ham}

In Sec. \ref{sec-gtr}, we obtained the gravitational Lagrangian
density as a function of $N$, $N_{\mu }$ and $h_{\mu \nu }$ as

\begin{equation}
{\cal L}(N, N_{\mu }, h_{\mu \nu }) = - \frac {\sqrt {h}N}{16\pi G}(
K^{2}-K_{\mu \nu }K^{\mu \nu }-^{3}\! R).
\end{equation}
The extrinsic curvature $K_{\mu \nu }$ involves time derivatives of
$h_{\mu \nu }$ and spatial derivatives of  $N_{\mu }$. The
three-curvature $^{3}\! R$ involves only spatial derivatives of $h_{\mu
\nu }$. Since the Lagrangian density does not contain time
derivatives of $N$ or $N_{\mu }$, the momenta conjugate to $N$ and
$N_{\mu }$ vanish:
 
\begin{equation}
\pi \equiv \frac {\delta {\cal L}}{\delta \dot {N}}=0,
\end{equation}

\begin{equation}
\pi ^{\mu }\equiv \frac {\delta {\cal L}}{\delta \dot {N}_{\mu }}=0.
\end{equation} 
These expressions are called primary constraints. The momenta
conjugate to $h_{\mu \nu }$ are

\begin{equation}
\pi ^{\mu \nu }\equiv \frac {\delta {\cal L}}{\delta \dot{h}_{\mu \nu
}} = \frac {\sqrt {h}}{16\pi G}(K^{\mu \nu } - h^{\mu \nu }K).
\end{equation} 
The gravitational canonical Hamiltonian for a closed geometry can now
be formed as 

\begin{equation}
{\cal H}_{c} = \int (\pi ^{\mu \nu }\dot {h}_{\mu \nu }+\pi ^{\mu }\dot
{N}_{\mu } +\pi \dot {N}-{\cal L})\; d^{3}x.
\end{equation} 

As usual for the Hamiltonian theory, one removes $\dot {h}_{\mu \nu
}$, $\dot {N}_{\mu }$ and $\dot{N}$ and express ${\cal H}_{c}$ in terms of the
coordinates $N$, $N^{\mu }$ and $h^{\mu \nu }$ and the conjugate
momenta $\pi $, $\pi ^{\mu }$ and $\pi ^{\mu \nu }$. Since the
primary constraints $\pi = \pi^{\mu }=0$ hold at all times, we have
 $\dot {\pi} =\dot { \pi} ^{\mu }=0$. Writing the Poisson brackets
for $\dot {\pi }$ and $\dot {\pi }^{\mu }$, we find

\begin{equation}
\dot {\pi }= \{ {\cal H}_{c},\pi \}=\frac {\delta {\cal H}_{c}}{\delta N}=0,
\label{eq:second1} 
\end{equation} 

\begin{equation}
\dot {\pi }^{\mu }= \{ {\cal H}_{c},\pi ^{\mu }\} = \frac {\delta {\cal H}_{c}}{\delta
N_{\mu }}=0. \label{eq:second2}
\end{equation} 
When generalised to include the matter variables and their conjugate
momenta, these expressions give the secondary constraints, which are
formally equivalent, respectively, to the time-time and time-space
components of the classical Einstein field equations.

The arena in which the classical dynamics takes place is called
`superspace', the space of all three-metrics and the matter field
configurations on a three-surface \cite{kolbturner}. This involves an
infinite number of degrees of freedom and hence to make the problem
tractable, all but a finite number of degrees of freedom must be
frozen out. The resulting finite dimensional superspace is known as a
`minisuperspace'. In the following, we consider a minisuperspace
model in which the only degrees of freedom  are those of the scale
factor $a$ of a closed RW spacetime and a spatially homogeneous
scalar field $\phi $. The Lagrangian for this problem is given by
(\ref{eq:ielphi}) with $k=+1$; i.e., 

\begin{equation}
L=-\frac {3\pi }{4G}N\left[ \frac {a\dot {a}^{2}}{N^{2}}-a-\frac
{8\pi G}{3}\left( \frac {\dot {\phi }^{2}}{2N^{2}}-V(\phi )\right)
a^{3}\right], 
\end{equation} 
from which we find the conjugate momenta $\pi _{a}$ and $\pi _{\phi
}$ as

\begin{equation}
\pi _{a}=\frac {\partial L}{\partial \dot {a}} = -\frac {3\pi }{2G}
\frac {a \dot {a}}{N} \label{eq:pia}
\end{equation} 
and 

\begin{equation}
\pi _{\phi }= \frac {\partial L}{\partial \dot {\phi }} = 2\pi
^{2}\frac {a^{3}\dot {\phi }}{N}.
\end{equation} 
The canonical Hamiltonian can now be constructed as 

\begin{eqnarray}
{\cal H}_{c}& = &\pi _{a}\dot {a}+ \pi _{\phi }\dot {\phi } -L \nonumber \\
&=& N\left[ -\frac {G}{3\pi }\frac {\pi _{a}^{2}}{a}-\frac {3\pi
}{4G} a+\frac {3\pi }{4G}a^{3}\left( \frac {G}{3\pi ^{3}}\frac {\pi _{\phi
}^{2}}{a^{6}} +\frac {8\pi G}{3} V(\phi )\right) \right] \nonumber \\
&\equiv & N{\cal H}.
\end{eqnarray} 
The secondary constraint (\ref{eq:second1}) now give 

\begin{equation}
{\cal H}=  -\frac {G}{3\pi }\frac {\pi _{a}^{2}}{a}-\frac {3\pi
}{4G} a+\frac {3\pi }{4G}a^{3}\left( \frac {G}{3\pi ^{3}}\frac {\pi _{\phi
}^{2}}{a^{6}} +\frac {8\pi G}{3} V(\phi )\right) =0, \label{eq:hamphi}
\end{equation} 
which is equivalent to (\ref{eq:t-tphi}). For the RW spacetime which
contains only a cosmological constant, (\ref{eq:iellambda}) helps us
to write the constraint equation  as

\begin{equation}
{\cal H}=  -\frac {G}{3\pi }\frac {\pi _{a}^{2}}{a}-\frac {3\pi
}{4G} a+\frac {3\pi }{4G}a^{3} 
\frac {8\pi G}{3} \rho _{\lambda } =0. \label{eq:haml}
\end{equation} 

This equation is equivalent to (\ref{eq:t-tdS}).  In all cases, ${\cal H}$
is independent of the lapse $N$ and shift $N^{\mu }$ and thus the
latter quantities are Lagrange multipliers (as mentioned towards the
end of Sec. \ref{sec-gtr}) and not dynamical variables.  Stated in a
different way, the fact that ${\cal H}=0$ is a consequence of a new symmetry
of the theory, namely, time reparametrisation invariance. This means
that using a new time variable $t^{\prime }$ such that $dt^{\prime }=
N\;dt$ will not affect the equations of motion. Also this enables one
to choose some convenient gauge for $N$, a procedure we adopt on
several occasions.  The constraint equation gives the evolution of
the true dynamical variable $h_{\mu
\nu }$ ($a$ in the above examples) and can be used in place of the
Hamilton equations.

\section{Wheeler-DeWitt Equation}
\label{sec-wd}

Canonical quantisation of  a classical system like the one above
means introduction of a wave function $\Psi (h_{\mu \nu }, \phi )$
\cite{kolbturner,halli,vilenkin} 
and requiring that it satisfies 

\begin{equation}
i\frac {\partial \Psi }{\partial t} = {\cal H}_{c}\Psi = N{\cal H}\Psi. \label{eq:gwd}
\end{equation}
To ensure that time reparametrisation invariance is not lost at the
quantum level, the conventional practice is to ask that the wave
function is annihilated by the operator version of ${\cal H}$; i.e.,

\begin{equation}
{\cal H}\Psi =0. \label{eq:wd}
\end{equation} 
But some other authors \cite{padijmp} argue that by defining
a new variable $\tau $ 
such that $N\;dt=d\tau $, one can retain the form $(\ref{eq:gwd})$;
i.e., 

\begin{equation}
i\frac {\partial \Psi }{\partial \tau }={\cal H}\Psi 
\end{equation}
and the resulting quantum theory will still be reparametrisation
invariant. However, in the following we use the more conventional form
(\ref{eq:wd}), which is called the Wheeler-DeWitt (WD) equation. 

This equation is analogous to a zero energy Schrodinger equation, in
which the dynamical variables $h_{\mu \nu },\; \phi $ etc. and their
conjugate momenta $\pi _{\mu \nu },\; \pi _{\phi }$ etc. (generally
denoted as $q^{\alpha }$ and $ p^{\alpha }$, in the respective order)
 are replaced by the corresponding operators. The wave
function $\Psi $ is defined on the superspace and we expect it to
provide information regarding the evolution of the universe. An
intriguing fact here is that the wave function is independent of
time; they are stationary solutions in the superspace. 
The wave functions commonly arising in quantum cosmology are of WKB
form and may be broadly classified as oscillatory, of the form
$e^{iS}$ or exponential, of the form $e^{-I}$. The oscillatory wave
function predicts a strong correlation between $q^{\alpha }$ and
$p^{\alpha }$ in the form

\begin{equation}
p_{\alpha } = \frac {\partial S}{\partial q^{\alpha }}.
\end{equation}
$S$ is generally a solution to the Hamilton-Jacobi equations. Thus
the wave function of the form $e^{iS}$ is normally thought of as
being peaked about a set of solutions to the classical equations and
hence predicts classical behaviour. A wave function of the form
$e^{-I}$ predicts no correlation between coordinates and momenta and
so cannot correspond to classical behaviour.

In a minisuperspace, one would expect $\Psi $ to be strongly peaked
around the  trajectories identified by the classical solutions. But
these solutions are subject to observational verification, at least
in the late universe so that a subset of the general solution can be
chosen as describing the late universe. Now the question is whether
the solution to the WD equation can discern this subset too. But it
shall be noted that, just like the Schrodinger equation, the WD
equation merely evolves the wave function and there are many
solutions to it. To pick one solution, the normal practice is to
specify the initial quantum state (boundary condition). These
boundary conditions, through the wave function, therefore set initial
conditions for the  solution of classical equations.  Then one may
ask whether or not the finer details of the universe we observe today
are consequences of the chosen theory of initial conditions.

In the simple example of the RW spacetime which contain only a
cosmological constant, $\Psi $ is defined on the minisuperspace with
one dimension in the variable $a$. We replace $\pi _{a} \rightarrow
i\;d/da $ in (\ref{eq:haml}) to write the WD equation  as

\begin{equation}
\left[ \frac {d^{2}}{da^{2}} - \frac {9\pi ^{2}}{4G^{2}}\left( a^{2}
-\frac {8\pi G}{3}\rho _{\lambda }a^{4}\right) \right]\Psi (a) =0 .
\label{eq:wdl}
\end{equation} 
The factor ordering in the operator replacement in (\ref{eq:haml}) is
ambigous. For many choices of factor ordering, the effect 
 can be parametrised by a constant $r$ and the corresponding
Hamiltonian operator is obtained by the substitution 

\begin{equation}
\pi _{a}^{2} \rightarrow -a^{-r}\left( \frac {\partial }{\partial
a}a^{r}\frac {\partial }{\partial a} \right). \label{eq:fo}
\end{equation}
The choice in (\ref{eq:wdl}) corresponds to $r=0$.  But  it will not
significantly affect the semiclassical calculations and hence we
choose the form in (\ref{eq:wdl}) for convenience. In this form the
WD equation resembles a one-dimensional Schrodinger equation written
for a particle with zero total energy, moving in a potential

\begin{equation}
U(a)=\frac {9\pi^{2}}{4G^{2}}\left( a^{2}-\frac {8\pi G}{3}a^{4}\rho
_{\lambda } \right) .\label{eq:ua}
\end{equation} 
Let us now define 

\begin{equation}
a_{0}\equiv \left( \frac {8\pi G}{3}\rho _{\lambda }\right)^{-1/2}.
\label{eq:a0} 
\end{equation}
In the particle analogy, there is a forbidden region for the zero
energy particle in the intervel $0<a< a_{0}$ and  a classically allowed
region for $a>a_{0}$. The WKB solutions of  (\ref{eq:wdl}) in the
classically allowed region $a>a_{0}$ are 

\begin{equation}
\Psi _{\pm }(a) = \pi _{a}^{-1/2} \exp \left[ \pm i \int
_{a_{0}}^{a}\pi _{a^{\prime }}da^{\prime }\mp i\pi /4 \right],
\end{equation} 
where $\pi _{a} = [-U(a)]^{1/2}$. In the forbidden region the solutions
are 

\begin{equation}
\bar {\Psi }_{\pm }(a) = \mid \pi _{a}\mid ^{-1/2}\exp \left[ \pm \int
_{a}^{a_{0}} \mid \pi _{a^{\prime }} \mid da^{\prime }\right] .
\end{equation} 
For $a\gg a_{0}$, we have

\begin{equation}
-i\frac {d}{da}\Psi _{\pm }(a) \approx \pm \pi _{a} \Psi _{\pm }(a).
\end{equation} 

Thus $\Psi _{-}$ and $\Psi _{+}$ describe, respectively,  an
expanding and contracting universe. It is now that we impose boundary
conditions and different boundary conditions lead to different
predictions. Some of such well-motivated proposals for the boundary
conditions are by Hartle-Hawking, Vilenkin and Linde
\cite{hh,lindeboundary,vilenkinboundary}. 

\section{Boundary Condition Proposals}
\label{sec-boundary}

The Hartle-Hawking `no boundary' boundary condition \cite{hh} is
expressed in terms of a Euclidean path integral. The corresponding
wave functon, in the present case, is specified by requiring that it
is given by $\exp (-I_{E})$ in the  under barrier regime, where
$I_{E}$ is the Euclidean action. This gives

\begin{equation}
\Psi _{H} (a<a_{0})= \bar {\Psi }_{-}(a),
\end{equation} 

\begin{equation}
\Psi _{H}(a>a_{0})=\Psi _{+}(a) - \Psi _{-}(a).
\end{equation} 

This corresponds to a real wave function with equal mixture of
expanding and contracting solutions in the classically allowed
region. Linde's wave function \cite{lindeboundary} is obtained by
reversing the sign of the exponential in the Euclidean regime;

\begin{equation}
\Psi _{L} (a<a_{0})= \bar {\Psi }_{+}(a),
\end{equation} 

\begin{equation}
\Psi _{L}(a>a_{0})=\frac {1}{2}\left[ \Psi _{+}(a) + \Psi _{-}(a)\right].
\end{equation} 
Vilenkin's `tunneling boundary condition' \cite{vilenkinboundary}
gives a purely expanding solution for the classical regime

\begin{equation}
\Psi _{T}(a>a_{0}) = \Psi _{-}(a)
\end{equation} 
and the under-barrier wave function is 

\begin{equation}
\Psi _{T}(a<a_{0}) = \bar {\Psi }_{+}(a) -\frac {i}{2}\bar {\Psi }_{-}(a).
\end{equation} 
The growing exponential $\bar {\Psi }_{-}(a)$ and the decreasing
exponential $\bar {\Psi }_{+}(a)$ have comparable amplitudes at
$a=a_{0}$, but away from that point the decreasing exponential
dominates. This, he describes as creation of the universe from
`nothing'. 

This quantisation scheme is applied to spacetimes
which contain scalar fields. The attempt is to examine the
possibility of emergence of a semiclassical phase from the quantum
cosmological era, which contains a scalar field with the required initial
conditions for inflation to occur.  On using the Hartle-Hawking wave
function, the probability for tunneling from $a=0$ to $a=a_{0}$ is
given by

\begin{equation}
P_{H} \propto e^{-I_{E}}.
\end{equation} 
Under the Vilenkin tunneling boundary condition,
 
\begin{equation}
P_{L} \propto e^{-\mid I_{E}\mid }.
\end{equation} 
If the potential of the scalar field has several extrema, then using
the latter prescription, tunneling favours the maximum with largest
value of $V(\phi )$ (which is advantageous for inflation) whereas the
former prescription favours the minimum with the smallest value of the
potential. However, all these authors agree that these proposals may
be criticised on the grounds of lack of generality or lack of
precision \cite{halli,vilenkin}.

\section{Quantisation of the New Physical Models}
\label{sec-newpq}

Quantisation of the models discussed in Ch. 3 involves a slight
paradigm shift: we do not have inflation and also our models always 
contain matter along with vacuum energy. Though our prototype model
is the one with zero energy-momentum tensor and  a complex
scale factor, we postpone the discussion on that to the next section.
First we consider our coasting model discussed in Sec.
\ref{sec-alter}, with total energy density varying as $a^{-2}$.

We adopt the approach of  Fil'chenkov \cite{fil}, who has considered the WD
equation for flat, closed and open universes which allow for some
kind of matter other than vacuum. He generalises the potential $U(a)$
given by (\ref{eq:ua}) for $\rho _{\lambda }$= constant by
writing the  energy density for the universe in the form

\begin{equation}
\tilde {\rho }=\rho _{pl}\sum _{n=0}^{6} B_{n}\left( \frac
{l_{pl}}{a}\right)^{n} .
\end{equation}
Here $n=3(1+w)$.This is a superposition of partial energy densities
of various kinds of matter at Plankian densities, each one of them
being separately conserved. The kinds of matter
included are

\begin{eqnarray}
n&=& 0 \qquad (w=-1) \qquad \hbox {vacuum}, \nonumber \\
n&=& 1 \qquad (w=-2/3) \qquad  \hbox {domain walls}, \nonumber \\
n&=& 2 \qquad (w=-1/3) \qquad \hbox {strings}, \nonumber \\
n&=& 3 \qquad (w=0) \qquad  \hbox {dust}, \nonumber \\
n&=& 4 \qquad (w=1/3) \qquad  \hbox {relativistic matter}, \nonumber \\
n&=& 5 \qquad (w=2/3) \qquad  \hbox {bosons and fermions}, \nonumber \\
n&=& 6 \qquad (w=1) \qquad  \hbox {ultra stiff matter}, \nonumber \\
\end{eqnarray}
The WD equation is now written as 

\begin{equation}
\left[ \frac {d^{2}}{da^{2}} - U(a) \right] \Psi =0,
\end{equation}
with the generalised form of the potential (\ref{eq:ua}) 

\begin{equation}
U(a) = \frac {9\pi ^{2}}{4G^{2}} \left( ka^{2} -\frac {8\pi G}{3}
a^{4}\tilde {\rho }\right) .
\end{equation}

We too proceed along similar lines, but first considering only a single
conserved component at a time.  It shall be noted that the constraint
(\ref{eq:t-t}) and field equations (\ref{eq:s-s}) for an energy
density $\rho = C_{n}/a^{n}$ with equation of state (\ref{eq:es})
(where $w=\frac {n}{3}-1$) are obtainable from the Lagrangian 

\begin{equation}
L=2\pi ^{2}a^{3}N\left[ -\frac {1}{16\pi G}\left( \frac {6}{N^{2}}\frac {\dot
{a}^{2}}{a^{2}} -\frac {6k}{a^{2}}\right) -\frac {C_{n}}{a^{n}}\right]
\end{equation} 
by writing the Euler-Lagrange equation corresponding to variation
with respect to $N$ and $a$. The Hamiltonian is

\begin{equation}
{\cal H}=  -\frac {G}{3\pi }\frac {\pi _{a}^{2}}{a}-\frac {3\pi
}{4G}k a+\frac {3\pi }{4G}a^{3} 
\frac {8\pi G}{3} \frac {C_{n}}{a^{n}} =0 \label{eq:hamc}
\end{equation} 
and the WD equation in this case can be written as

\begin{equation}
\left[ \frac {d^{2}}{da^{2}} - \frac {9\pi ^{2}}{4G^{2}}\left(k a^{2}
-\frac {8\pi G}{3}a^{4-n}C_{n}\right) \right]\Psi _{n}(a) =0. \label{eq:wdc} 
\end{equation} 
For $n>2$, classically there is a forbidden region for $a>a_{0}$,
whereas the allowed region is for $a<a_{0}$; $a_{0}\equiv [(8\pi G/3)
C_{n}]^{1/(n-2)}$. 
We see that for the special case with $n=2$, the WD equation reduces to

\begin{equation}
\left[ \frac {d^{2}}{da^{2}} - \frac {9\pi ^{2}}{4G^{2}}a^{2} \left(
k-\frac {8\pi G}{3}C_{2}\right)\right]\Psi =0.
\end{equation} 
With $C_{2}=(3/8\pi G)(m^{2}+k)$, this corresponds to the energy
density (\ref{eq:rhotchen2}) advocated 
by us. In
this case, the WD equation is simply 

\begin{equation}
\left[ \frac {d^{2}}{da^{2}}+\frac {9\pi ^{2}}{4G^{2}}a^{2}m^{2}\right]
\Psi =0.
\end{equation} 
It is clear that $\Psi $ is oscillatory for all values of $a$.  If we
choose the factor ordering corresponding to $r=-1$ [instead of $r=0$:
See Eq. (\ref{eq:fo})] in the above, we have

\begin{equation}
\left[ \frac {d^{2}}{da^{2}}- \frac {1}{a}\frac {d}{da}+\frac {9\pi
^{2}}{4G^{2}}a^{2}m^{2}\right] 
\Psi =0,
\end{equation} 
which has an exact solution

\begin{equation}
\Psi (a) \propto \exp \left( {\pm i \frac {3\pi }{4G} ma^{2}}\right).
\label{eq:psic}
\end{equation} 
It is of interest to note that if we define $\Psi \equiv e^{iS}$ in
the above, $S$ satisfies the Hamilton-Jacobi equation

\begin{equation}
\left( \frac {dS}{da}\right)^{2} +U(a) =0,
\end{equation}
where $U(a) = -(9\pi ^{2}/4G^{2})a^{2}m^{2}$. The classical
constraint in this case is $\pi _{a}^{2} +U(a) =0$. This invites the
identification 

\begin{equation}
\pi _{a}^{2} = \left(\frac {dS}{da}\right)^{2} = \frac {9\pi
^{2}}{4G^{2}} a^{2}m^{2}.
\end{equation}
Using this in our definition  $\pi _{a} \equiv \partial L/\partial
\dot{a} = -(3\pi /2G)a \dot {a}$ [as in (\ref {eq:pia}), with $N=1$],
we get $\dot {a} =\pm m$, from which the coasting evolution is
regained. Thus the oscillatory wave function 
 is strongly
peaked about the singular coasting evolution throughout the history
of the universe.

Now let us turn to the physical universe with total energy density
$\tilde {\rho }$ given by (\ref{eq:rhot}).
Clearly the field equation and constraint (\ref{eq:fe}) follow from
the Lagrangian 

\begin{equation}
L=\frac {3\pi }{4G}\left( -\frac {\dot {a}^{2} a}{N^{2}}-a+\frac
{a_{0}^{2}}{a}\right) .\label{eq:lagp}
\end{equation}
The Hamiltonian is 

\begin{equation}
{\cal H}=-\frac {G}{3\pi }\frac {\pi _{a}^{2}}{a} + \frac {3\pi }{4G}\left(
a- \frac {a_{0}^{2}}{a} \right)=0,
\end{equation}
so that the WD equation, with factor ordering $r=0$, is

\begin{equation}
\left[ \frac {d^{2}}{da^{2}} - \frac {9\pi ^{2}}{4G^{2}} \left(
a_{0}^{2}-a^{2}\right) \right]\Psi =0. \label{wdp}
\end{equation} 
The potential in this case indicates that $a<a_{0}$ is a classically
forbidden region.
The classical action $\int L\;dt$ constructed using (\ref{eq:lagp})
in this under-barrier region  can be seen to be 

\begin{equation}
S=i\frac {3\pi }{2G}a_{0}^{2}\{ \frac {1}{2}\cos ^{-1}\left( \frac
{a}{a_{0}}\right) -\frac {1}{4} \sin 2\left[ \cos ^{-1} \left( \frac
{a}{a_{0}} \right) \right] \} ,
\end{equation}
which is pure imaginary. It can be seen that for $a\ll a_{0}$,
$e^{iS}$ satisfies the WD equation. Similarly for the region
$a>a_{0}$, the classical action is evaluated as

\begin{equation}
S= \frac {3\pi }{2G} \left[ \left( a^{2}-a_{0}^{2}\right) ^{1/2} a-
a_{0}^{2}\cosh ^{-1}\left( \frac {a}{a_{0}} \right) \right]
\end{equation}
which is real. Also in this case, $e^{iS}$ is a solution for $a\gg
a_{0}$. Using a reasoning like that in the case of (\ref{eq:psic}),
we can regain the solution (\ref{eq:aotc})  in both cases.

\section{Quantisation of the Complex, Source-free Model}
\label{sec-complexq}

Lastly, we quantise the model with complex scale factor and zero
energy-momentum tensor \cite{jj2} and show that this model has the correct
classical correspondence with the classical trajectory. From
(\ref{eq:cact}), the Lagrangian for the problem is obtained as $L=
-(3\pi /4G) \left( \dot {\hat {a}}^{2} \hat {a} - \hat {a}\right) $.
The conjugate momentum to $ \hat {a} $ is

\begin{equation}
\pi _{ \hat {a}} = \frac { \partial L }{ \partial \dot {\hat {a}}} 
=-\frac {3 \pi}{2G} \hat{a} \dot {\hat {a}}.
\end{equation}
The  Hamiltonian is 

\begin{equation}
{\cal H} = -\frac {G}{3 \pi} \frac {\pi _{\hat {a}}^{2}}{\hat {a}}- 
\frac {3\pi }{4G} \hat {a}.
\end{equation}
The constraint equation ${\cal H} = 0 $ has the corresponding WD equation 

\begin{equation}
({\cal H}- \epsilon) \Psi (\hat {a})= 0 \label{eq:wdepsilon}
\end{equation}
where we have made a modification such that an arbitrary real
constant $\epsilon $ is introduced to take account of a possible energy
renormalisation in passing from the classical constraint to its
quantum operator form, as done by Hartle and Hawking in \cite{hh}.
It shall be noted that this equation is still reparametrisation
invariant. Choosing the operator ordering for the sake of simplicity
of the solution, we get,

\begin{equation}
\frac {d^{2} \Psi (\hat{a})}{d \hat {a}^{2}} - (\frac {9\pi
^{2}}{4G^{2}} \hat {a}^{2} +\frac {3\pi }{G} \epsilon \hat {a} ) \Psi
(\hat {a}) = 0
\end{equation}
 Making  a substitution $ \hat {S}
= \sqrt {3\pi /2G} \left[ \hat {a} + (2G/3\pi ) \epsilon \right] $,
this becomes, 

\begin{equation}
\frac {d^{2} \Psi (\hat {S})}{d \hat {S} ^{2}} + (\frac 
{2G}{3\pi }   \epsilon ^{2} -  \hat {S}^{2}) \Psi (\hat {S}) = 0.
\end{equation}
The wave equation has ground state harmonic oscillator type solution for 
$\epsilon  =\sqrt { 3\pi /2G} $:

\begin{equation}
\Psi (\hat {a} ) = {\cal {N}}\; \; \exp  \left[ -\frac 
{3\pi}{4G}\left( \hat {a} + \sqrt {\frac {2G}{3\pi }} \right)
^{2}\right]. \label{eq:psicom}
\end{equation}
This is nonnormalisable, but it is not normal in quantum cosmology to 
require that the wave function should be normalised \cite{halli}.
Our choice is further justified by noting that the probability density

\begin{equation}
\Psi ^{\star }\Psi = {\cal {N}}^{2}\;\; \exp  \left( \frac {3\pi
}{2G}y^{2} \right) \; 
\exp \left[ -\frac {3\pi }{2G}\left( x + \sqrt {\frac {2G}{3\pi }}
\right) ^{2}\right]  
\end{equation}
is sharply peaked about the classical contour  given by Eq.
(\ref{eq:ac2}), which is a straight line parallel to the imaginary
axis with $ x$ remaining a constant. We can identify $a_{0}$ with the
expectation value of $x$;

\begin{equation}
a_{0} \equiv <x> =  - \sqrt {\frac {2G}{3\pi }}  \label{eq:a0c}
\end{equation}
so that $\mid a_{0} \mid \approx l_{pl} $, which is the desired
result.  The $\exp \left[ \left( 3\pi /4G\right) y^{2}\right] $ part
of the wave function is characteristic of a Riemannian space-time with
signature (+ + + + ).  This is precisely the feature we should expect
to correspond to the imaginary part in the scale factor.

The classical correspondence can be made more explicit by making an
argument similar to that made in Sec. \ref{sec-newpq}. The classical
Hamiltonian constraint equation in this case is 

\begin{equation}
\pi _{\hat {a}}^{2} + \frac {9\pi ^{2}}{4G^{2}}\hat {a}^{2} =0.
\end{equation}
Defining the above wave function (\ref{eq:psicom}) as $\Psi (\hat
{a})\equiv e^{iS}$, we note that $S$ satisfies the Hamilton-Jacobi
equation 

\begin{equation}
\left( \frac {dS}{da}\right)^{2} + \frac {9\pi ^{2}}{4G^{2}}\left(
\hat {a}+ \sqrt {\frac {2G}{3\pi }}\right) ^{2}=0
\end{equation}
Comparing these two equations, we see that $e^{iS}$ is strongly
peaked about the classical solution, for large $\hat {a}$ when
compared to the Planck length. 

Thus the result obtained on quantisation is that the simplest minimum
energy wave function is sharply peaked about the classical contour of
evolution of $\hat {a}$, just like the ground state harmonic
oscillator wave function in quantum mechanics is peaked about the
classical position of the particle. But we welcome the important
difference with this analogy; ie., the quantum mechanical system in
our case is not localised. In fact, the wave function is not
normalisable along the imaginary axis. If it was with real scale
factor, the exponential growth of the wave function would correspond
to some classically forbidden region, but in this case, we have the
nonnormalisable part for the wave function along the imaginary axis;
this result is just what we should expect since it corresponds to our
classical system and cannot be termed as `classically forbidden'. The
most significant fact is that the quantum cosmological treatment
helps us to predict the value of $a_{0}$, the minimum radius in the
nonsingular model as compared to the Planck length.

\chapter{Reprise}

\section{Comparison of Solutions}

For the purpose of comparison with the solution of Einstein equations
in the new cosmological model, we take a closer look at the occurrence
of inflation mentioned in Sec. \ref{sec-inflation}. In the scalar
field model described by (\ref{eq:t-tphi}) -(\ref{eq:consphi}), we
see that the field equations (\ref{eq:s-sphi}) and (\ref{eq:consphi})
are second order partial differential equations, whose solution
involves initial values of four quantities $a$, $\dot {a}$, $\phi $
and $\dot {\phi }$. The constraint equation (\ref{eq:t-tphi})
connects first derivatives and hence the number of arbitrary
parameters in the theory gets reduced to three.  The occurrence of
inflation in this system is not generic; it depends crucially on
several factors \cite{halli}.  First of all, the potential $V(\phi )$
should be of the inflationary type; i.e., that for some range of
values of $\phi $, $V(\phi )$ should be large and $\mid \frac
{dV(\phi )}{d\phi }/V(\phi ) \mid \ll 1$. For the subset of $k=+1$
solutions, inflation occurs only when the initial value of $\dot
{\phi } \approx 0$. It is argued that quantum cosmology provides such
initial conditions favourable for inflation to occur, by the choice
of proper boundary conditions. In this case,  the cosmological wave
function is peaked around the trajectories defined by

\begin{equation}
\dot {a} \approx \left[ \frac {8\pi G}{3} a^{2}V(\phi )\right] ^{1/2}
\gg 1, \qquad \dot {\phi }\approx 0.
\end{equation}
Then the number of free parameters are reduced to two. Integrating
the above equations give

\begin{equation}
a\approx \exp \left[ \sqrt {\frac {8\pi G}{3}}V^{1/2}(t-t_{0})\right]
, \qquad \phi \approx \phi _{0} = \hbox {constant}
\end{equation}
Here $t_{0}$ and $\phi _{0}$ are the two arbitrary constants
parametrising this set of solutions. The constant $t_{0}$ is in fact
irrelevant, because it is the origin of the unobservable parameter
time. However, this solution is  inflationary.

Let us contrast this situation with the solution of the system
described by (\ref{eq:t-tc}) and (\ref{eq:s-sc}). The complex field
equation (\ref{eq:s-sc}) is in fact a set of two second order partial
differential equations and involves four free parameters. But also
the constraint equation contains two first order equations:

\begin{equation}
\dot {x}=0, \qquad \dot {y}=\pm 1
\end{equation}
This helps us to obtain the desired solution  $\hat {a} =a_{0}\pm
it$, which corresponds to the nonsingular physical model, as a
general one and without resorting to quantum cosmology. Since $t_{0}$
is irrelevant, $a_{0}$ is effectively the only free parameter in the
classical theory.  Quantum cosmology, in fact, allows us to predict
this value too, as comparable to Planck length. This prediction is
not in the way $a_{0}$ is identified in the conventional quantum
cosmological theories mentioned in Sec. \ref{sec-boundary} or in our
own models discussed in Sec. \ref{sec-newpq}. In these cases, $a_{0}$
can be readoff from the potential itself and is not obtained as an
expectation value using the wave function.

Another feature that distinguishes our quantum cosmological treatment
is that we are not imposing any adhoc boundary conditions; we only
look for an exact solution to the WD equation. This procedure is
quite similar to the solution of harmonic oscillator problem in
ordinary quantum mechanics. In this sense, introduction of a zero
point energy in the WD equation (\ref{eq:wdepsilon}) is justifiable.
However, we adopt the point of view that the vanishing of the
classical Hamiltonian  can be taken care of by restricting the
solution to that corresponding to the minimum energy. It is of
interest to note that this minimum (zero point) energy is $\epsilon
=(3\pi /2)^{1/2}\epsilon _{pl}$ where $\epsilon _{pl}$ is the Planck
energy. That is, the total energy of the universe is not zero; it is
the Planck energy - apart from a numerical factor.

At this point, it is worth while to point out that  the total
positive energy (matter, vacuum etc.) contained in the closed real
universe at $t=0$, evaluated using (\ref{eq:rhot}) and (\ref{eq:a0c})
is also equal to $\epsilon $. The negative energy contributes a value
$-\epsilon
/2$ so that the total energy is $\epsilon /2$. (This, of course, does
not include the gravitational energy).
 
Coming back to the solution of the complex field equation
(\ref{eq:s-sc}), we may now state why we assumed $k=+1$ and did not
consider the $k=0$ and $k=-1$ cases. It can be seen that if we
require the complex spacetime and the corresponding physical universe
to have the same value of $k$, then the $k=0,\; -1$ cases are unsuitable to
describe the universe we livein. For $k=0$, the constraint equation
gives $\dot {x}=0$, $ \dot {y}=0$, which leads to a static universe.
For $k=-1$, it is true that we get a solution $\hat {a}=\pm
t+ia_{0}$ from which $\mid \hat {a}\mid ^{2} = a_{0}^{2} + t^{2}$ is
obtainable. But in this case, the physical universe has to obey the
equation

\begin{equation}
\frac {\dot {a}^{2}}{a^{2}} - \frac {1}{a^{2}} = -\frac
{a_{0}^{2}}{a^{4}} ;
\end{equation}
i.e., the physical universe contain only the negative energy density.
For these reasons, we do not consider these two possibilities as
viable and set $k=+1$ at the outset.

\section{Coasting Evolution}

The physical models we obtain in both approaches (Secs. \ref{sec-new},
\ref{sec-alter}) have coasting evolution. In the latter model, it is
coasting throughout the history and in the former, it coasts when the
universe is a few Planck times or more old. Historically, the first
coasting cosmological model is the Milne universe. To understand
this model, first consider a two dimensional flat spacetime given in
coordinates $(t,X)$ with the metric $ds^{2} =dt^{2}-dX^{2}$. Let the
worldline $L_{0}$ be the line $X=0$. By repeatedly using the Lorentz
boost corresponding to some small velocity $\Delta V_{0}$, a family
of worldlines which all pass through $O$ can be generated. A model in
which these are the worldlines of fundamental observers represents an
expanding universe obeying the cosmological principle. All the
fundamental observers are equivalent to each other and because the
worldline $L_{0}$ is a straight line representing inertial motion,
the same is true for other worldlines too. Since the Lorentz boost is
repeated infinitely often, an infinite number of worldlines are
obtained by this construction. A four dimensional analogue of this
model is usually referred to as Milne universe. This is a flat space
cosmological model, not incorporating the effects of
gravitation \cite{ruthellis,sahnimilne}. 
Alternatively, it is described by a flat, empty spacetime having a RW
metric with $k=-1$, $a(t)=t$ and $q=0$.

Coasting cosmologies are encountered in many situations including
non-Einstein theories of gravitation (\cite {sahnimilne} and references
therein. In the Friedmann models itself, it is easy to see from
(\ref{eq:ddota}) that coasting evolution results when $\rho +3p =0$,
our models being examples. The quantity $\rho +3p$ is sometimes
referred to as gravitational charge. An interesting property of such
spacetimes was recently pointed out by Dadhich {\sl et al.}
\cite{dadhich}. They 
resolve the Riemann tensor, which characterises the gravitational
field into electric and magnetic parts, in analogy with the
resolution of the electromagnetic field. It can be seen  that the
electric part is caused by mass-energy while the magnetic part is due to
motion of the source. But unlike other fields, gravitation has two
kinds of charges; one is the usual mass-energy and the other is the
gravitational field energy. Consequently, also the electric part is
decomposed into an active part, which is Coulombic and a passive part,
which produces space curvature. An interchange of active and passive
electric parts in the Einstein equation, which is referred to as
electrogravity duality transformation, is shown to be equivalent to
the interchange of Ricci and Einstein curvatures. These authors show
that under this transformation, spacetimes with $\rho+3p=0$ go over
to flat spacetime; i.e., they are dual to each other.

Absence of a particle horizon, agreement with the predicted age of
the universe etc. in a coasting evolution are well known, but since
it is usually considered as a feature of spacetimes containing only
some exotic matter like strings, textures etc., this most simple
cosmological scenario is not given serious attention in the
literature. The Ozer-Taha model is coasting, but only upto the
relativistic era and deviates from it after that epoch. Our physical
model demonstrates that a coasting evolution throughout the history
of the universe is a  promising contender to a realistic
cosmological model, which resolves all outstanding problems in the
standard cosmology and at the same time  not making  too drastic
modifications to it.

\section{Avoidance of Singularity}

The physical model obtained in Sec.  \ref{sec-real} is a bounce solution from a
previous collapse, rather than an explosion from a big bang singularity.  Such a
bounce is sometimes referred to as a `Tolman wormhole' \cite{molina,coule}.
Oscillating universes have somewhat similar features and were considered as
alternatives to the big bang cosmologies in the earlier literature, but interest
in such cyclical evolution declined after the first cosmological singularity
theorems.  Recently, the quasi-steady state theory \cite{qss} revives this
scenario.  An analysis of bounce solutions reveals that the absolute minimum
requirement for this to occur is the violation of (only) the strong energy
condition (SEC).  The various energy conditions, in the context of Friedmann
models are the following \cite{molina}:

\begin{eqnarray}
\hbox {Null energy condition (NEC)} &\Leftrightarrow & \rho +p \geq 0
\nonumber \\
\hbox {Weak energy condition (WEC)} &\Leftrightarrow & \rho \geq 0
\; \; \hbox {and} \; \; \rho +p\geq 0 \nonumber \\
\hbox {Strong energy condition (SEC)} &\Leftrightarrow & \rho +3p\geq 0
\; \; \hbox {and} \; \; \rho +p \geq 0 \nonumber \\
\hbox {Dominant energy condition (DEC)} &\Leftrightarrow  & \rho \geq 0
\; \; \hbox {and} \; \; \rho \pm p \geq 0 \nonumber \\
\end{eqnarray}

It is shown that in a $k=+1$ universe, only the SEC need to be violated for
obtaining a bounce solution.  Since the singularity theorems mentioned above use
the SEC as an input hypothesis, violating this condition vitiates them \cite
{visser}.  Physically, violating the other energy conditions with (small)
quantum effects is relatively difficult.  On the other hand, it is rather easy
to violate the SEC and is therefore often referred to as `the unphysical energy
condition'.  Using $\tilde {\rho }$ and $\tilde {p}$ given by (\ref{eq:rhot})
and (\ref{eq:pt}) in the above energy conditions, we can see that our
nonsingular model satisfies all the energy conditions except the strong one and
serves as a perfect example for this phenomenon.

When comparing our two physical models, it is clear that the avoidance of
singularity is primarily due to the presence of the negative energy density.
The naturalness of a negative energy density at the classical level may be
suspect.  But we should note that the nonzero value of $a_{0}$ on which this
depends is obtained on quantisation.  However, as mentioned before, negative
energy densities were postulated much earlier.  Currently, there is a revival of
interest in negative energies in connection with speculations on wormholes,
time-travel etc.  \cite{morris,visser}.  Also some speculations are on which
consider a Casimir driven evolution of the universe \cite{antonsen}.  That
negative energy densities are predicted by relativistic quantum field theory is
known for a long time.  Casimir \cite{casimir}, for the first time, showed that
between two parallel perfect plane conductors separated by a distance $l$, there
is a renormalised energy $E= -\pi ^{2}/720 l^{3}$ per unit area and this is now
experimentally confirmed.  The energy density corresponding to this may be
evaluated as $-\pi ^{2}/720 l^{4}$.  The Casimir energy density is calculated
for some static universe models.  For example, this density for a massless
scalar field in the four-dimensional static Einstein universe is \cite
{elizalde}

$$
\rho _{Casimir} = -\frac {0.411505}{4\pi ^{2}a^{4}}
$$
A similar expression for an expanding closed universe is not known to
us. However, we shall compare the above value with our expression for
negative energy density (\ref{eq:rho-}), with $a_{0}$ given by
(\ref{eq:a0c}): i.e.,

$$
\rho _{-} = -\frac {1}{4\pi ^{2}a^{4}}.
$$
Anyhow, it will be premature to identify $\rho_{-}$ with Casimir
energy, just like identifying $\rho _{\lambda }$ with vacuum energy.

\section{Prospects and Challenges}

In this subsection, we discuss some of the possible future
developments  in connection with the new
model, both observational and conceptual.

1) Consider the physical nonsingular model with real scale factor
$a(t)=(a_{0}^{2} +t^{2})^{1/2}$. This model is obtainable directly
from the assumption that the universe is closed and has a total
energy density and pressure given by Eqs. (\ref{eq:rhot}) and
(\ref{eq:pt}) respectively. The assumption of complex scale factor
etc. serves the purpose of justifying this one. It is shown that
globally, the model has very good predictions and is devoid of all
the cosmological problems mentioned in Sec. \ref{sec-problems}. But
to be compatible with modern observational cosmology, it has
to go a long way. Of utmost importance is the fluctuations in CMBR
detected by COBE; any realistic cosmological model should be able to
account for this. In Sec. \ref{sec-thermal}, it was argued that the
present model can generate density perturbations on scales as large
as the present Hubble radius, even after the nucleosynthesis epoch.
Recently, Coble {\sl et al.} \cite{coble} have claimed that while
models with a constant' cosmological constant have too high a COBE
normalised amplitude for a scale invariant spectrum, their
decaying-$\lambda $ model has this amplitude matching with
observations. However, a detailed analysis of CMBR
anisotropies is not undertaken here. Another issue of importance which we have
not looked into in any detail is the
nucleosynthesis in the present model. It is shown that the thermal
histories of the new model and the standard model are not very different.
Hence it would  be reasonable to expect that nucleosynthesis will
also proceed identically.

2) If the standard model is to be generalised by including some kind of
energy density other than relativistic/nonrelativistic matter, the
resulting model cannot remain unambitious for long; it invariably has to
get connected to field theory or the `standard model' in particle
physics. In that sense, the present model has only put forward a
phenomenological law for the evolution of the vacuum energy which we
prudently call $\lambda $ (or $\rho _{\lambda }$). A field theoretic
explanation for $\rho _{\lambda }$ will always be welcome. In fact,
one can see some resemblance between the set of equations
(\ref{eq:t-tbeta})-(\ref{eq:betadot}) and
(\ref{eq:t-tphi})-(\ref{eq:consphi}), which suggests the possibility
of considering $\beta $ as a field. It is easy to see that this is
not an ordinary scalar field; it is more akin to a Brans-Dicke field.
This aspect too is not pursued any further.

3) Another important issue worthy of further exploration is the
connection with quantum stationary geometries' (QSG's)
\cite{qsg,pad}. As an example, this theory juxtaposes two situations;
one in which a classical system of closed dust filled universe  with
constraint equation 

\begin{equation}
\frac {\dot {a}^{2}}{a^{2}} + \frac {1}{a^{2}} = \frac {A}{a^{3}}
\end{equation}
having a singular evolution for the scale factor $a=a_{class.}(t)$
and the other in which QSG's avoid this singularity in such a way
that  

\begin{equation}
<a^{2}(t)> = a_{0}^{2} + a_{class.}^{2}(t)
\end{equation}
Also here, $a_{0}$ is shown to be of the order of Planck length. This
is analogous to the avoidance of singularity in the new and its
alternative models. This and many other aspects of the quantum
behaviour in the model are left untouched.

\medskip

 Lastly, some aspects of aesthetics.  It is well known that Einstein considered
the right hand side of his equation, which contain a nongeometric quantity (the
energy-momentum tensor) as spoiling the consistency and integrity of his
geometrical approach.  In the present case, we do not hesitate to claim that at
least in a cosmological context, a realistic model is obtained in which such a
voluntary introduction of a nongeometrical quantity is not necessary.  In fact,
equations (\ref{eq:t-tbeta})-(\ref{eq:betadot}) are essentially the same
equations (\ref{eq:t-tc}) and (\ref{eq:s-sc}) and hence it can be considered
that the right hand sides of (\ref{eq:t-tbeta})-(\ref{eq:s-sbeta}) or that of
(\ref{eq:constrainte})-(\ref{eq:fe}) as emerging from their corresponding left
hand sides.

One cannot simply be averse to the philosophical overtones of
this theory. The universe with complex scale factor is the
unperceived one, but the same field equations describe a real,
physical universe with real scale factor. Our intellect can conceive
only the measurable, real quantities and in a sense, this makes the
energy-momentum tensor nonzero. If not approached with caution, this
can lead to mysticism, but perhaps it would be better to interpret
this, in the event of being  proved to have some truth content,  as
yet another instance in physics where, to use N. Bohr's words,
"truths being statements in which the opposite also are truths".

This position can be criticised on two grounds. (1) The observational
and theoretical uncertainties are greatly amplified in cosmology and
hence it is subjected to all sorts of ideological and philosophical
influences, the present theory being one example. But it shall be
reminded that none of the existing cosmological models are free from
it and at the level of analysis made, the present model has equally
good, if not better, predictions. (2) At a  subtler level, it can be
argued that it is our intellect that imposes its laws upon nature. We
quote K. Popper \cite{popper}, who remarked on this subject in reply
to Kant: "Kant was right that it is our intellect which imposes its
laws - its ideas, its values - upon the inarticulate mass of  our
"sensations" and thereby brings order into them. Where he was wrong
is that he did not see that we rarely succeed with our imposition,
that we try again and again, and that the result - our knowledge of
the world - owes as much to the resisting reality as to our self
produced ideas".  We note that this makes the task of conforming to
any epistemological systematics difficult for the scientist.

As a closing note, we remark that the model with complex scale factor
can be considered  as a model for an underlying objective reality.
The theory is clearly falsifiable; in the predictions $H_{p}t_{p}=1$,
$q_{p}=0$, $\rho _{m}/\rho _{\lambda }=2$ in the nonrelativistic era, 
the total energy density $\tilde {\rho }$, the negative energy density
$\rho _{-}$
etc., it leaves no adjustable parameters.  Though  it looks a
mathematical curiosity, at best a toy model, it is curious enough how
this simple model can account for this much cosmological observations
without creating any problems at a physical level.

\end{document}